\documentclass{aa}  
\usepackage{graphicx}
\usepackage{txfonts}
\usepackage{hyperref}
\usepackage{natbib}
\defcitealias{EHT1}{EHT L1}
\defcitealias{EHT3}{EHT L3}
\defcitealias{EHT4}{EHT L4}
\defcitealias{EHT5}{EHT L5}
\defcitealias{EHT6}{EHT L6}
\defcitealias{porth19}{Porth et al. 2019}

\usepackage{url}
\usepackage{color}
\usepackage{xcolor}
\usepackage{xspace}
\usepackage[normalem]{ulem}

\newcommand{\gy}{\textsc{Gyoto}\xspace}

\newcommand{\be}{\begin{equation}}
\newcommand{\ee}{\end{equation}}
\newcommand{\bea}{\begin{eqnarray}}
\newcommand{\eea}{\end{eqnarray}}
\newcommand{\nn}{\nonumber}
\newcommand{\pp}{\varphi}

\graphicspath{{Figure/}}

\begin{document}

\title{Geometric modeling of M87* as a Kerr black hole \\ or a non-Kerr {compact} object}

\author{
  F. H. Vincent\inst{1}
  \and
  M. Wielgus\inst{2,3}
  \and
  M. A. Abramowicz\inst{4,5,6}
  \and
  E. Gourgoulhon\inst{7}
  \and
  J.-P. Lasota\inst{8,4}
  \and
  T. Paumard\inst{1}
  \and
  G. Perrin\inst{1}
}

\institute{
  LESIA, Observatoire de Paris, Universit\'e PSL, CNRS, Sorbonne Universit\'es, UPMC Univ. Paris 06, Univ. de Paris, Sorbonne Paris Cit\'e, 5 place Jules Janssen, 92195 Meudon, France
  \email{frederic.vincent@obspm.fr}
  \and
  Black Hole Initiative at Harvard University, 20 Garden St., Cambridge, MA 02138, USA
  \email{maciek.wielgus@gmail.com}
  \and
  Center for Astrophysics $|$ Harvard \& Smithsonian, 60 Garden Street, Cambridge, MA 02138, USA
  \and
  Nicolaus Copernicus Astronomical Center, Polish Academy of Sciences, Bartycka 18, PL-00-716 Warszawa, Poland
  \and
  Physics Department, University of Gothenburg, 412-96 G{\"o}teborg, Sweden
  \and
  Institute of Physics, Silesian University in Opava, Czech Republic
  \and
  Laboratoire  Univers  et  Th\'eories,  Observatoire  de  Paris,  Universit\'e PSL, CNRS, Universit\'e de Paris, 92190 Meudon, France
  \and
  Institut d'Astrophysique de Paris, CNRS et Sorbonne Universit\'e, UMR 7095, 98bis Bd Arago, F-75014 Paris, France
}

% \abstract{}{}{}{}{} 
% 5 {} token are mandatory

\abstract
{The Event Horizon Telescope (EHT) collaboration
  recently obtained first images of the surroundings
  of the supermassive compact object M87* at the center of
  the galaxy M87. This provides a fascinating probe of the properties
  of matter and radiation in strong gravitational fields.
  It is important to determine from the analysis of these results, what can and what cannot
  be inferred about the nature of the spacetime around M87*}
{We want to develop a simple analytic disk model for the accretion
  flow of M87*. 
  Compared to
  general-relativistic magnetohydrodynamic (GRMHD) models, it has the
  advantage of 
  being independent of the turbulent character of the flow, and controlled by only few
  easy-to-interpret, physically meaningful
  parameters.
  We want to use this model to predict the image of M87*
  assuming that it is either a Kerr black hole, or an alternative compact object.}
{We compute the synchrotron emission from the disk model
  and propagate the resulting light rays to the far-away
  observer by means of relativistic ray tracing. Such computations
  are performed assuming different spacetimes (Kerr, Minkowski, non-rotating ultracompact star,
  rotating boson star or Lamy spinning wormhole). We perform numerical fits of these models to the EHT data.}
{We discuss the highly-lensed features of Kerr images and show that they are intrinsically linked to the accretion-flow properties, and not only to gravitation. This fact is illustrated by the notion of \textit{secondary ring} that we introduce. 
  Our model of spinning Kerr black hole predicts mass and orientation consistent with the EHT interpretation.
  The non-Kerr images result in similar quality of the numerical fits and may appear very similar to Kerr images, once blurred to the EHT resolution. This implies
  that a strong test of the Kerr spacetime 
  may be out of reach
  with the current data. We notice that future developments of the EHT could alter this situation.}
{Our results show the importance of studying alternatives
  to the Kerr spacetime in order to be able to test the
  Kerr paradigm unambiguously. More sophisticated treatments
  of non-Kerr spacetimes, and more advanced observations, are needed in order to go further
  in this direction.}

\keywords{Physical data and processes: Gravitation -- Accretion, accretion discs -- Black hole physics -- Relativistic processes -- Galaxies: individual: M87
}

\maketitle
%
%-------------------------------------------------------------------

\section{Introduction}
The galaxy Messier 87 (M87) is a giant elliptical galaxy
located in the Virgo cluster,
first observed by the French astronomer Charles Messier in 1781.
Since a century it has been known to give rise to a kiloparsec-scale
radio jet~\citep{curtis18}.
The central engine of this jet is likely a supermassive
black hole, M87*. It is, like our Galactic Center, a low-luminosity
galactic nucleus, displaying a hot, optically thin {and most likely geometrically thick} accretion/ejection
flow~\citep{yuan14}.
The distance to M87 is of the order of the mean distance to
the Virgo cluster, that is $16.5$~Mpc~\citep{mei07}.
The mass of M87*
has been assessed to be $3.5\times 10^9$~M$_\odot$ by means
of gas-dynamics fitting~\citep{walsh13} and to $6.6\times10^9$~M$_\odot$
by means of stellar-dynamics study~\citep{gebhardt11}.

The Event Horizon Telescope (EHT) collaboration has recently
published the first reconstructed millimeter images of the close vicinity
of M87*~\citepalias{EHT1}. The images show a circular crescent 
feature with a diameter of $\approx 40~\mu$as,
with a non-isotropic flux distribution, surrounding
a central fainter region. These features are in good
agreement with what is known from theoretical imaging
of black holes~\citep{bardeen73,luminet79,marck96,ck2015,cunha18}.
The crescent morphology of the source was constrained by "free-form" imaging~\citepalias{EHT4}, {simple} geometric models, as well as direct fitting to GRMHD simulations~\citepalias{EHT6}. The large collection of GRMHD simulations created a framework for the physical interpretation of the EHT results \citepalias{EHT5,porth19}.
This analysis allowed to interpret the $40~\mu$as circular
feature as a lensed accretion/ejection flow
within a few $M$ from the black hole. The non isotropy
can be linked to a relativistic beaming effect.
The central fainter region is consistent with being
the shadow of the black hole \citep{falcke2000}. Within this framework,
the mass of M87* was estimated to be $6.5\pm0.7 \times 10^9~$M$_\odot$,
assuming a distance of $16.8\pm0.8$~Mpc \citepalias{EHT1}, which is in agreement with the independent stellar dynamics measurement. 
For the images shown in this article, we use the consistent values of $6.2 \times 10^9~$M$_\odot$ for the mass and $16.9$~Mpc for the distance, following the choice
made in~\citetalias{EHT5}.

The assumptions of the GRMHD-based analysis and interpretation have given
rise to theoretical investigations regarding the nature of
the features seen in the EHT images~\citep{gralla19,Johnson2019,narayan19,gralla19b}. 
The main question is to what extent these features can
be directly linked to gravitation, and how much are they influenced by the highly
model-dependent astrophysics of the emission. There are at least several effects to consider in this context, corresponding to particular choices and simplifications made in the EHT's GRMHD simulations library~\citep[][and references therein]{porth19}. Those include, but are not limited to utilizing a prescription for electron temperature, ignoring the dynamical feedback of radiation, viscosity, resistivity and the presence of non-thermal electrons \citepalias{EHT5}. Apart from that, the turbulent character of the flow adds time dependence to the model, with {poorly understood, possibly strongly resolution--dependent, relationship between the simulations and real variability of the source \citep[see, e.g.,][]{White2019}}. Given all those uncertainties, it is both interesting and important to interpret the EHT measurements in the framework of simple physically motivated geometric models. So far, such models have not been extensively discussed in the context of the M87* image interpretation. Only \citet{nalewajko19} have recently adopted a geometric model, but with a simple powerlaw prescription for the emission,
and no absorption.

{
  Models for M87* environment have been published,
  using both analytical or GRMHD descriptions of the
  flow. Analytical models used disk-dominated
  radiatively inefficient accretion flows (RIAF)
  or RIAF+jet models~\citep{yuan00,diMatteo03,broderick09b}.
  GRMHD models are describing the disk+jet environment
  of M87*~\citep{dexter12,moscibrodzka16,davelaar19}.
}

The aim of this paper is to contribute to the physical interpretation
of the EHT images by using a simple {analytical} geometric
model that is able to capture the most prominent features
of a more realistic setup,
avoiding the uncertain astrophysics embedded in the latter.
For simplicity, we restrict ourselves here to a pure
disk model, not taking into account any ejection feature.
{We stress that the origin of the photons forming the
   EHT image might be the base of the M87 jet, or the disk.
   This point was investigated in~\citetalias{EHT5} with a library of state-of-the-art GRMHD simulations. Among a variety of models, only the SANE models, with particularly high $R_\mathrm{high}$ parameter~\citep{moscibrodzka16}, were found to be dominated by the jet emission at that scale. For most of the models, including all MAD ones, the emission observed by the EHT is actually dominated by the disk component.}
%\sout{This setup can describe, e.g., a magnetically arrested state of accretion%, where the innermost disk
%  emission dominates over the jet emission contribution}
%~\citepalias[][note that emission from the low-density, strongly-magnetized inner part of the jet funnel is masked for all simulations in the EHT GRMHD library]{EHT5}.
We consider thermal synchrotron emission and
absorption in this disk.
{Our geometric model of the plasma surrounding the black hole is
  as simple as possible. This simplicity allows to be as little sensitive as
  possible to the uncertainties that affect more elaborate models.
  We believe that such a framework is well adapted for testing the
  impact of the central compact object's gravitation on the observables.
  }
Our goals are to (1) discuss the prominent Kerr image features obtained within this context with a particular emphasis on the accretion-model dependence of the highly-lensed regions, and (2) 
{try to answer the question whether the EHT images, analyzed 
independently of a broader astrophysical context and of external constraints, can deliver a test of the Kerr-spacetime paradigm. For that purpose we compare the accretion disk images computed for several different models of spacetime.} While certain non-Kerr spacetimes were briefly discussed in \citetalias{EHT1} and \citetalias{EHT5}, all quantitative considerations by the EHT consortium were performed within the framework of the Kerr spacetime paradigm. We aim to fill this gap with the current paper.
We highlight that throughout this article, we always consider that gravitation is described by
general relativity. While we consider different spacetimes, that may require exotic form of the stress-energy tensor, the Einstein's theory of gravitation
is never modified. 

This paper is organized as follows: section~\ref{sec:kerr} considers that M87*
is a Kerr black hole.
After introducing our disk model in section~\ref{sec:diskmodel},
we present millimeter-wave Kerr images in section~\ref{sec:kerrimages}, where we discuss the origins of these images main features, related to the properties of
highly-bent null geodesics.
Section~\ref{sec:nonaxisym} briefly discusses the modification
in the image generated by non-axisymmetric structures.
Section~\ref{sec:nonkerr} is dedicated
to studying how the M87* image changes when the spacetime
is different from Kerr.
We consider the Minkowski spacetime (section~\ref{sec:minko}),
the spacetime of a static ultracompact star with an emitting surface
(section~\ref{sec:star}), 
a rotating boson--star spacetime (section~\ref{sec:BS}), and a
Lamy wormhole spacetime (section~\ref{sec:lamy}). 
In section~\ref{sec:EHTfit}, we discuss fits to the
EHT data of our Kerr and non-Kerr models.
Section~\ref{sec:conc} gives
conclusions and perspectives.

\section{Emission from a thick disk in a Kerr spacetime}
\label{sec:kerr}

In this section, the Kerr spacetime is labeled by means of the
Boyer-Lindquist spherical coordinates $(t,r,\theta,\pp)$.
We work in units where
the gravitational constant and the speed of light are
equal to $1$, $G = c = 1$. Radii are thus expressed in units
of the black hole mass $M$.

\subsection{Disk model and emission} 
\label{sec:diskmodel}

\begin{figure}[htbp]
\centering
\includegraphics[width=0.5\textwidth]{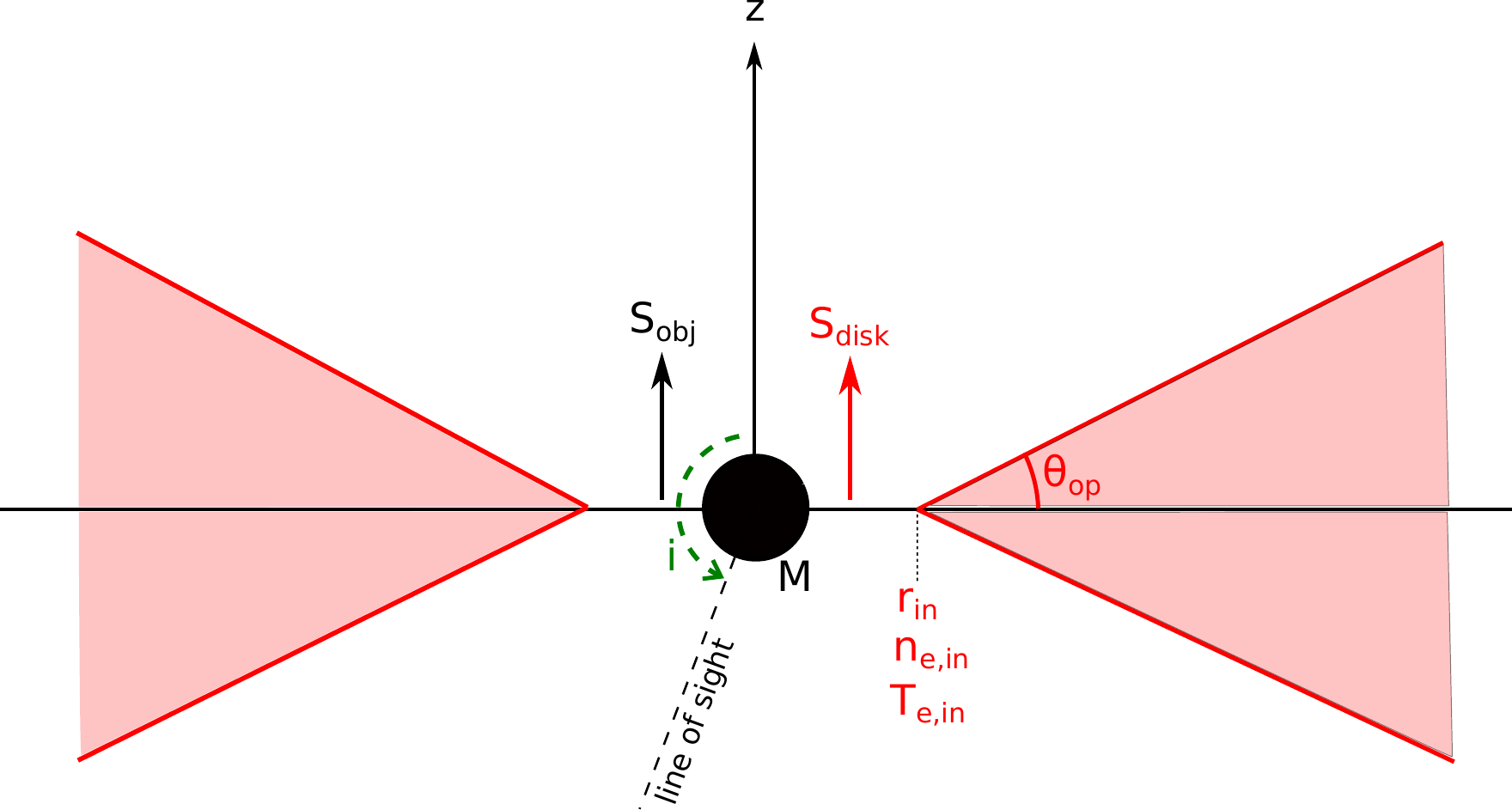}
\caption{Geometrically thick disk model (in red) surrounding
a compact object (black disk) of mass $M$. 
The disk has an inner
  radius $r_\mathrm{in}$, where the electron number density and
  temperature are $n_\mathrm{e,in}$ and $T_\mathrm{e,in}$. The number
  density scales as $r^{-2}$ and the temperature as $r^{-1}$.
  All quantities are independent of the height $z$.
  The opening angle of the disk is called $\theta_\mathrm{op}$. Here
  and in the remaining of the article, the compact object and
  accretion disk spins (see respectively the black and red arrows)
  are assumed to be aligned.
  The inclination angle $i$ between the spin axis
  and the line of sight is shown in green.
} 
\label{fig:disk}
\end{figure}

We consider a geometrically thick, optically thin accretion
disk in a setup illustrated in Fig.~\ref{fig:disk}.

For simplicity, we parametrize the geometry of the accretion
disk by only two parameters, its inner radius $r_\mathrm{in}$
and opening angle $\theta_\mathrm{op}$. We do not prescribe any outer radius
for the disk. It is effectively imposed by selecting a field of view for computing the images, as well as 
by the radially decaying profiles of temperature and density.
The disk is assumed to be axisymmetric with respect
to the $z$ axis which lies along the black hole spin.
The compact object and accretion-disk spins are assumed to be aligned, as is the case for the entire EHT GRMHD library.
{Throughout this article we 
fix the opening angle of the to $\theta_\mathrm{op} = 30^\circ$, so the disk is moderately geometrically thick. This choice places our considerations between the limit cases of geometrically thin model considered by \citet{gralla19} and spherical accretion considered by \citet{narayan19}, and within the thickness range expected for a real accretion flow in M87* \citep{yuan14}}.

We model the emission by thermal synchrotron radiation.
This is also a simplification because shocks and turbulence
in the accretion flow are likely to generate non-thermal emission.
However, we chose to neglect this additional complexity. This is
again primarily for the sake of simplicity, but also because
the broad features of the image are unlikely to be extremely
sensitive to the details of the emission process, and because
the non-thermal emission modeling would necessarily imply somewhat
arbitrary extra assumptions anyway. 
Thermal synchrotron emission is modeled following the formulas
derived by~\citet{pandya16}. {Both the emission and absorption
coefficients are self-consistently taken into account in our computation}. These coefficients depend on the electron number
density and temperature, as well as on the magnetic field strength.
 The parameters of our model are the density and temperature at the inner disk radius,
$n_\mathrm{e,in}$ and $T_\mathrm{e,in}$.
We assume simple power laws for their scaling, $n_\mathrm{e,in} \propto r^{-2}$ for the density, and $T_\mathrm{e,in} \propto r^{-1}$ for the temperature, following the description of~\citet{vincent19}.

We do not consider any vertical variation of
the accretion flow properties. As for the magnetic field prescription,
we simply impose the magnetization $\sigma = B^2 / (4\pi) / (m_p c^2 n_e)$,
equal to the ratio of the magnetic to particle energy densities,
with $B$ being the magnetic field magnitude, $m_p$ the proton mass,
and $c$ the velocity of light (kept here for clarity). This quantity is always set to $\sigma = 0.1$
in this article. 
This is an arbitrary choice, which has little impact
on our results given that we do not discuss a mixed disk+jet model, in which case the magnetization should typically differ in the
disk and in the jet.
{We note that our choice of power laws for the electron density,
temperature, and the magnetic field are that of the standard 
model of~\citet{blandford79}. It also agrees with the inner
evolution of these quantities in GRMHD simulations of M87*~\citep[see, e.g.,][]{davelaar19}. We stress that only the
inner few tens of $M$ of the flow matters for the images that we
discuss here. We thus parametrize the radial dependence of the
disk quantities in order to capture the relevant properties of this region.}

For all the images shown in this article,
we assume the observing frequency of
$\nu_\mathrm{obs,0} = 230$~GHz, corresponding to the observing frequency of the EHT.
The orientation of the model is determined by the assumption that the jet aligns with the black hole\,/\,disk spin axis and by the observed jet position angle on the sky. We fix the inclination (angle between the black hole spin
and the line of sight) to $i=160^\circ$ \citep{Walker2018},
meaning that the black-hole and disk spin vectors
are directed "into the page" for all images
presented here \citepalias{EHT5}, see Fig. \ref{fig:disk}.
{This nearly face-on inclination may be in general unfavorable for considering deviations in the image geometry, when compared to near-edge-on
  views~\citep{bardeen73}.}
The position angle
of the approaching jet (angle east of north of the black hole spin projection
onto the observer's screen plane) is fixed to $PA=290^\circ \equiv -70^\circ$ \citep{kim2018}.
The field of view of the presented images is fixed to $f=160\,\mu$as and the number of pixels to $200 \times 200$ (unless otherwise noted).

One extra crucial assumption has to be made: the choice of the
dynamics of the accretion flow. We will always consider Keplerian
rotation outside
of the innermost stable circular orbit (ISCO), irrespective of
the height $z$ with respect to the equatorial plane.
If the inner radius
is smaller than $r_\mathrm{ISCO}$, the emitting matter 4-velocity
below ISCO is given as
\be
\mathbf{u}_\mathrm{em} = \Gamma \, (\mathbf{u}_\mathrm{ZAMO} + \mathbf{V}) \  ,
\ee
where $\mathbf{u}_\mathrm{ZAMO}$ is the 4-velocity of the
zero-angular-momentum observer (ZAMO) and $\mathbf{V}$
is the accretion flow velocity as measured by the ZAMO.
It can be written as
\be
\mathbf{V} = V^r \frac{\boldsymbol{\partial}_r}{\sqrt{g_{rr}}} +  V^\pp \frac{\boldsymbol{\partial}_\pp}{\sqrt{g_{\pp\pp}}} \ ,
\ee
so that $(V^r)^2 + (V^\pp)^2 = V^2 = (\Gamma^2 - 1) / \Gamma^2$.
This velocity is parametrized by choosing $V \in [0,1]$
and $v^\pp \equiv V^\pp/V \in [0,1]$. In the following, {we always fix $V$ to its value at the ISCO. For the two spin-parameter values considered, $a=0$ and $a=0.8M$,
this gives respectively $V=0.5$ and $V=0.61$}. We can then chose $v^\pp$ to simulate a limit case corresponding either to a flow with purely circular velocity (if $v^\pp=1$), or a radially plunging flow
(if $v^\pp=0$).
Note that the emitter velocity is always
independent of the height $z$, defined using Boyer-Lindquist coordinates by
$z = r\,\cos\theta$.

We ensure that the observed flux is of the order of $0.5 - 1$~Jy,
in agreement with the state of M87* at the time of the 2017 EHT campaign.
The flux is primarily impacted by the choice of the
electron number density and temperature at the inner radius.
Given that these two quantities are degenerate because we are
fitting a single flux value (rather than the full spectrum),
we decide to fix the electron number density at $r=2M$
to $n_\mathrm{e,2M} = 5 \times 10^5\,\mathrm{cm}^{-3}$, 
which is in reasonable agreement with the results published
in the literature by various authors~\citep[][\citetalias{EHT5}]{broderick09b,davelaar19}. The number density
at the chosen value of $r_\mathrm{in}$ is thus fixed by the assumed $r^{-2}$ density scaling. The inner temperature $T_\mathrm{e,in}$ is
then chosen to obtain a reasonable value of the observed flux.
We find that the choice of $T_\mathrm{e,in}=8\times 10^{10}$~K
(or $k T_\mathrm{e,in} / m_e c^2 = 13.5$ in units
of the electron rest mass) leads
to reasonable flux values for all setups considered here.

The final step of our simulation is to perform general-relativistic
ray tracing, either in the Kerr spacetime or in other geometries, to obtain theoretical images.
This is done using the open-source ray tracing code \gy~\citep[see][and \url{http://gyoto.obspm.fr}]{vincent11,vincent12} to
compute null geodesics backwards in time, from a distant
observer located at the distance of $D=16.9$~Mpc away from the disk.
We summarize the fixed properties of the model and images in Table~\ref{tab:properties}.
\renewcommand{\arraystretch}{1.2}
\begin{table}[h!]
\centering
 \begin{tabular}{c c c } 
 \hline
 Symbol &  Value & Property \\  %\vspace{0.5cm}
 \hline
 $M$ & 6.2 $\times 10^{9}$~M$_\odot$ & compact object mass  \\
  %\hline
 $D$ & 16.9~Mpc & compact object distance \\
 %\hline
 $\theta_\mathrm{op}$ & 30$^\circ$ & disk opening angle   \\ 
 %\hline
 $n_\mathrm{e,2M}$ & $5 \times 10^5\,\mathrm{cm}^{-3}$ & max number density of electrons  \\
 %\hline
 $T_\mathrm{e,in}$ & $8\times 10^{10}$~K & max electron temperature \\
 %\hline
 $\sigma$ & 0.1 & magnetization \\
 %\hline
 $i$ & 160$^\circ$ & inclination angle\\
  %\hline
 $PA$ & -70$^\circ$ & jet position angle east of north \\ %in the image \\
  %\hline
 $\nu_\mathrm{obs,0}$  & 230~GHz & observing frequency \\
   %\hline
 $f$ & 160~$\mu$as & field of view \\
    %\hline
 -- & 200$\times$200 & image resolution \\
 [1ex]
 \hline
\end{tabular}
\caption{Fixed properties of the M87* models and images assumed throughout this paper.}
\label{tab:properties}
\end{table}

\begin{figure*}[htbp]
\centering
\includegraphics[width=0.95\textwidth]{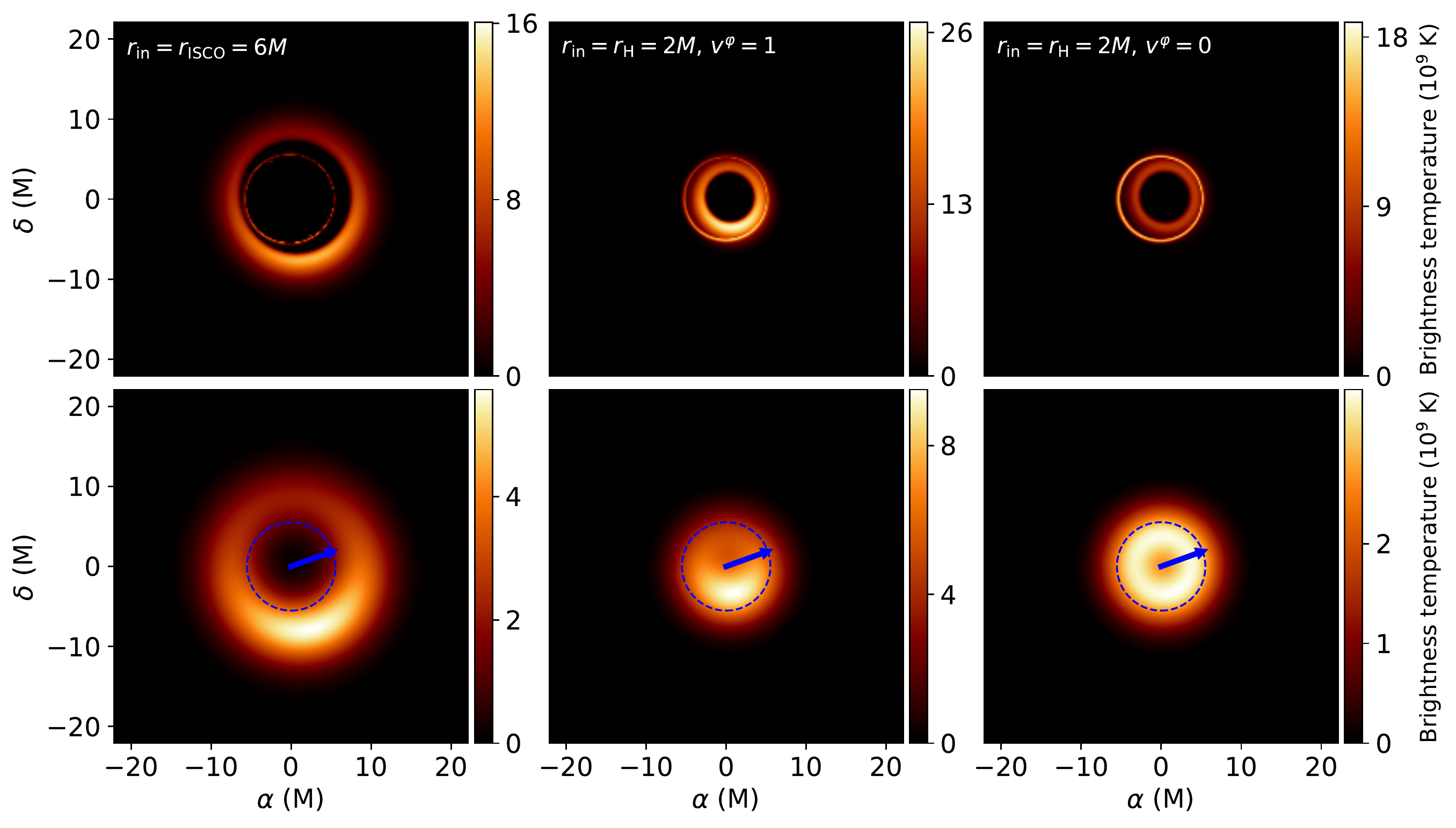}
\caption{Images
  of a thick disk surrounding a non-rotating ($a=0$)
  black hole. The top row shows the simulated ray-traced image,
  the bottom row consists of the upper row images blurred to the EHT resolution (about $20~\mu$as). The dashed blue circle shown in the lower row images has a diameter
  of $40\,\mu$as, consistent with the estimated diameter of the ring feature in the M87* image, reported by the EHT. This diameter translates to 11.05$M$ in mass units of distance. The blue arrow shows
  the projected direction of the jet.
  The disk inner radius is $6M$ for the left panel (corresponding
  to the ISCO) and $2M$ for the two other panels (corresponding
  to the event horizon). The azimuthal velocity below ISCO is
  paramterized by $v^\pp=1$ (purely azimuthal velocity)
  for the middle panel and $v^\pp=0$ (purely radial plunge)
  for the right panel. The inner electron number density
  is equal to $5\times10^5\,\mathrm{cm}^{-3}$ when $r_\mathrm{in}=2M$
  and $5.5\times10^4\,\mathrm{cm}^{-3}$ when $r_\mathrm{in}=6M$
  (see text for details on how these numbers are chosen). 
} 
\label{fig:disk_a0}
\end{figure*}
\begin{figure*}[htbp]
\centering
\includegraphics[width=0.95\textwidth]{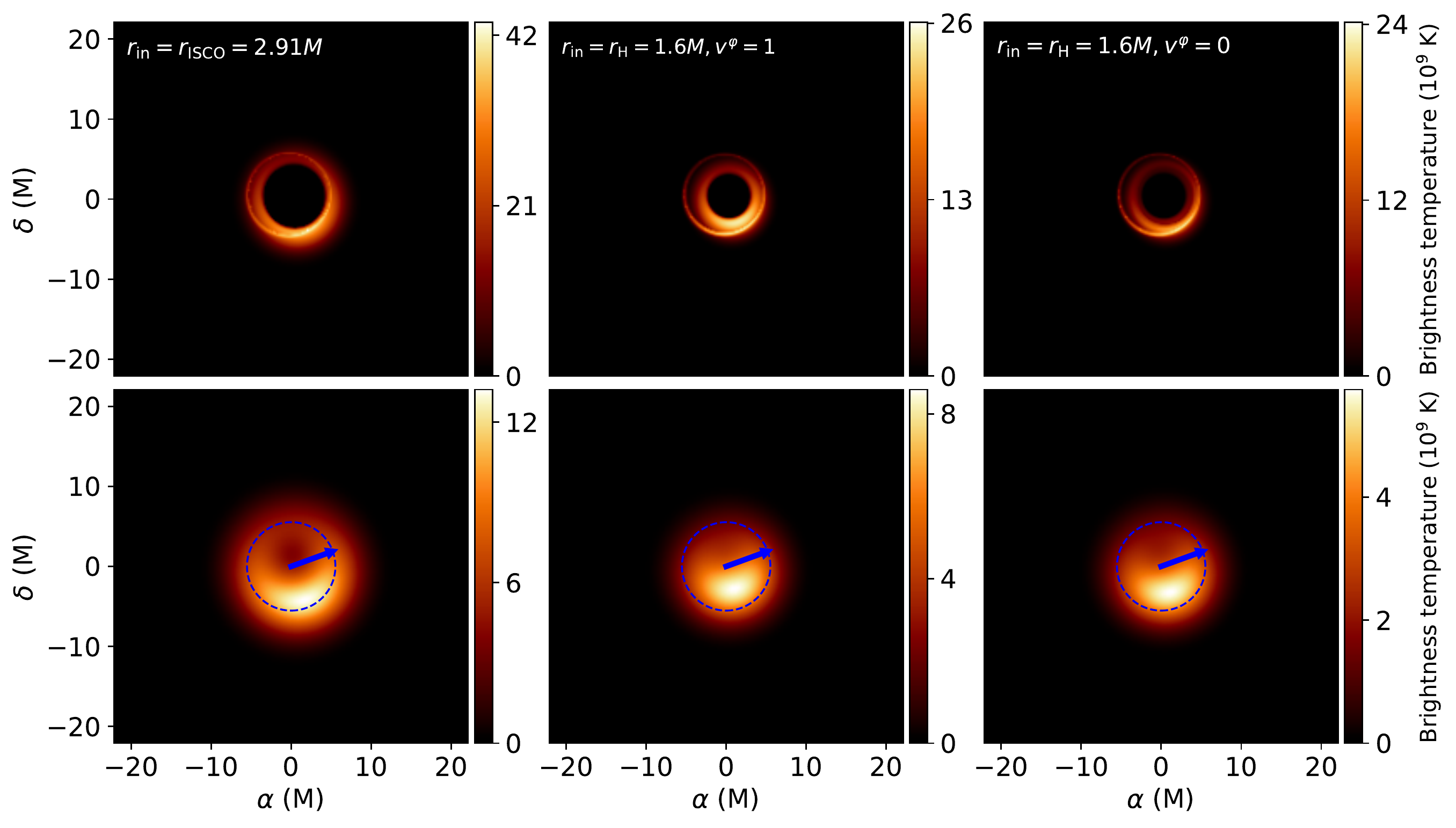}
\caption{Same as Fig.~\ref{fig:disk_a0} with a spin parameter
  of $a=0.8M$. The inner density is of
  $2.4\times10^5\,\mathrm{cm}^{-3}$ when $r_\mathrm{in}=2.91M$,
  and $7.8\times10^5\,\mathrm{cm}^{-3}$ when $r_\mathrm{in}=1.6M$.
  } 
\label{fig:disk_a08}
\end{figure*}

%-------------------------------
%-------------------------------
\subsection{Main features of the images}
\label{sec:kerrimages}
%-------------------------------
%-------------------------------

Figures~\ref{fig:disk_a0}-\ref{fig:disk_a08} show
the Kerr-spacetime disk images obtained for two different
values of the spin parameter. In this section, we discuss
the main features of these images, focusing
on the impact of the flow geometry and dynamics, as well
as on the highly-lensed-flux portion of the image, 
generally loosely called the "photon ring".

Figure~\ref{fig:disk_a0} shows the resulting image for a spin parameter
$a=0$ and three different choices for the accretion disk properties.
{We note that such non-rotating configurations are
unlikely to account for the {powerful large-scale jet of M87 \citepalias{EHT5}}. These configurations are still of interest for a comparison with
the rotating ones.} 
The top-left panel shows a Keplerian flow with inner radius at the Schwarzschild spacetime ISCO,
$r_\mathrm{ISCO}=6M$.
The top-middle panel has an inner radius going down to the
event horizon at $r_\mathrm{H}=2M$, with a purely azimuthal velocity below
the ISCO ($v^\pp=1$). The top-right panel is the same as the middle
panel, but with $v^\pp=0$ (pure radial inflow). The bottom panels show the same
images, convolved with a Gaussian kernel with full width at half maximum
of $20\,\mu$as, which is approximately the EHT angular resolution \citepalias{EHT4}. 
In this image, as well as in the following ones, we indicate the approximate position of the crescent feature reported by the EHT, 40~$\mu$as, with a dashed circle. For the compact object mass and distance assumed in this paper, this translates into $11.05M$ diameter.
Figure~\ref{fig:disk_a08} shows the same setup for a spin of $a=0.8M$,
with $r_\mathrm{ISCO}=2.91M$ and $r_\mathrm{H}=1.6M$.

The unblurred images presented in Figs.~\ref{fig:disk_a0} and~\ref{fig:disk_a08}
all show thick annular areas with the addition of a very thin
bright ring. The thick annular area is due to the emission from the inner parts of the disk, with a feeble lensing effect on the null geodesics.
It is generally referred to as the \textit{primary image}
of the disk, and is composed of geodesics that cross the equatorial plane at most once~\citep{luminet79}. The polar radius of the primary image 
clearly varies with the assumed $r_\mathrm{in}$. The brightness
distribution with azimuthal angle in the primary image
is a consequence of the special-relativistic
beaming effect: parts of the flow coming towards the observer are
boosted. This effect is clearly visible in the left and central upper panels of Fig.~\ref{fig:disk_a0} where the flow is in circular rotation and coming towards the observer
in the south direction. The upper-right panel of
the same figure is obtained 
when the inner radius is set at the event horizon and the
flow velocity is chosen to be purely radial below the ISCO.
In this case, the flux distribution is less dependent
on the azimuth than for the circularly-rotating cases. We have however checked that, when taking into account special-relativistic effects only, the image becomes boosted in the west direction where the flow approaches the observer. 
The blurred image of this
radial-inflow case (lower-right panels of
Figs.~\ref{fig:disk_a0})
is thus very isotropic, which is not consistent with
the observed EHT image. This elementary discussion shows
that the size of the primary image and its flux
distribution with azimuthal angle are directly linked to
the choice of the inner radius and to the dynamics of the gas
in the inner disk regions. This result agrees with that
obtained by~\citet{nalewajko19} with a simpler model.

The very thin bright ring, also present in the unblurred images,
is often loosely referred to as the photon ring, and considered
to be the image on sky of the unstable Kerr equatorial 
prograde photon orbit.
However, the set of orbits that actually matters in order to
form this highly lensed feature is the set of spherical
Kerr photon orbits first analyzed by~\citet{teo03}, with numerous
recent developments~\citep[see, e.g.,][]{cunha17,Johnson2019}.
These are bound {unstable} photon orbits evolving at constant Boyer-Lindquist
radii, with periodical excursion in the $\theta$ direction
(the span of this excursion, $\theta_\mathrm{min} < \theta < \theta_\mathrm{max}$,
depends on the photon's angular momentum).
{The orbits are not periodic in $\pp$ and are either prograde or retrograde, depending on the sign
of the photon's conserved angular momentum}.
These orbits exist within a radial range $r_\mathrm{ph,pro} < r < r_\mathrm{ph,retro}$,
where $r_\mathrm{ph,pro}$ and $r_\mathrm{ph,retro}$ are the usual
Kerr equatorial prograde and retrograde photon orbit radii.
In particular, for the Schwarzschild spacetime in which only
one photon orbit exists at $r_\mathrm{ph} = 3M$, the set of spherical
photon orbits is simply the sphere $r=3M$.
The thin bright ring in Kerr images is thus due to light
rays that approach a spherical Kerr photon orbit before
reaching the far-away observer. 

%\fv{Check carefully this reformulated paragraph}.
As stated above, the spherical Kerr photon orbits are
periodic in $\theta$. The complete $\theta$ excursion
from $\theta_\mathrm{min}$ to $\theta_\mathrm{max}$ (or the other way round) can be covered by a null geodesic an arbitrary number
of times $n$, corresponding to $n$ crossings of the equatorial plane, before leaving the orbit and reaching
the far-away observer (remember that these orbits are unstable). As $n$ increases, the Boyer-Lindquist radius {of such an orbit becomes very close to that of a spherical photon orbit
and the impact point on sky tends to the critical curve}. 
Thus, the thin bright ring is actually
the sum of an exponentially converging sequence of sub-rings lying
at smaller and smaller polar radii on sky,
{and labeled by the number $n$ of crossings
of the black hole equatorial plane (see Figure~\ref{fig:subrings_secondaryring})}. This fact was first noted by~\citet{luminet79} for the Schwarzschild case. The resolution
of the image truncates this sequence at a finite number
of sub-rings (see, e.g., the lower-right panel
of Fig.~\ref{fig:illustrate_secondary} where the 
outermost sub-ring is clearly seen, the subsequent 
sub-ring is only barely visible, and the following ones
are lost due to finite resolution).
Note that for the M87* image, the complete set of sub-rings of
the thin bright ring lies within $\lesssim 1\,\mu$as
on sky so that a very high resolution would be
needed to resolve some of its components.
\citet{gralla19} use the term \textit{lensing ring} for
the outermost such sub-ring 
(corresponding to the set of geodesics that cross the equatorial plane exactly twice), while they keep the terminology \textit{photon ring} for the sum of all
subsequents sub-rings (corresponding to the set of geodesics that cross the equatorial plane more than twice).

\citet{Johnson2019} give an analytic expression %of the
for the limiting curve on sky, towards which the
series of sub-rings converge in the limit 
of $n \rightarrow \infty$. 
This limiting curve was called
the \textit{critical curve} by~\citet{gralla19} and we keep
this name. 
Introducing $\xi$ -- the polar radius on the
observer's screen in units of $M$, and $\phi$ -- the polar angle
on the observer's screen, the critical curve reads
\bea
\label{eq:polarcritical}
\xi &=& \sqrt{a^2\left(\cos i - u_+ u_-\right) + \ell^2}, \\ \nn
\phi &=& \mathrm{arccos}\left(-\frac{\ell}{\xi\,\sin i} \right) ,\\ \nn
\eea
where
\bea
u_{\pm} &=& \frac{r}{a^2 (r-M)^2} \left[-r^3 + 3M^2r - 2a^2 M \right. \\ \nn
&& \left. \pm 2\sqrt{M (r^2 - 2Mr + a^2) (2r^3 - 3Mr^2 + a^2 M) } \right], \\ \nn
\ell &=& \frac{M(r^2 - a^2) - r (r^2 - 2Mr + a^2)}{a (r-M)}. \\ \nn
\eea
Here, $r$ is the Boyer-Lindquist radius of the Kerr spherical photon
orbit followed by the photon on its way to the observer.
One counter-intuitive property of this critical curve, 
already discussed by~\cite{Johnson2019}, is the fact that
one Kerr spherical photon orbit (one value of $r$) 
is mapped to 2 values of the polar angle, $\phi$ and $2 \pi - \phi$.
This is due to the arccos definition of $\phi$. As a consequence,
the critical curve should be seen as the image on the sky
of the set of Kerr spherical orbits, with each spherical orbit
being mapped to 2 points along the curve. Note that actually only
a subset of the full set of Kerr spherical photon orbits
($r_\mathrm{ph,pro} < r < r_\mathrm{ph,retro}$) is imaged
on the sky, depending on the value of the inclination $i$. Only
for $i=90^\circ$ does the full set get imaged on the sky~\citep[see Fig.~2 of][]{Johnson2019}.

\begin{figure*}[htbp] 
\centering
\includegraphics[width=0.85\textwidth]{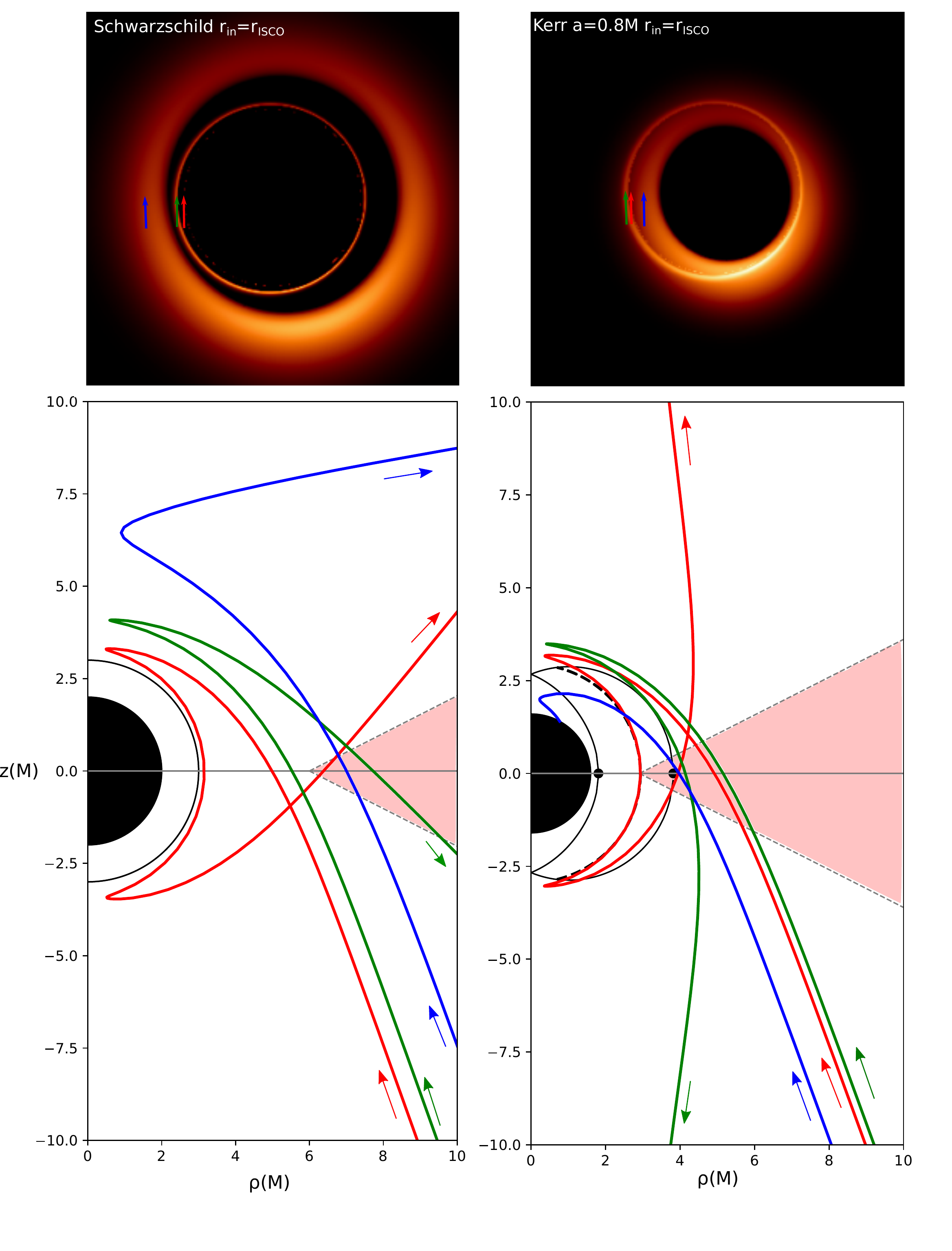}
\caption{\textit{Top panels:} zoom on the central $80~\mu$as
field of the image of a thick disk surrounding a Schwarzschild 
black hole (left) or a Kerr black hole with spin parameter $a=0.8M$ (right).
\textit{Lower panels}: three geodesics are plotted on the
$(\rho,z)$ plane of height vs cylindrical radius (in units of $M$). 
The arrows show the direction of backward-ray-tracing integration
in \gy. The observer is located at $16.9$~Mpc towards the lower right of the panels.
These geodesics correspond to the
pixels labeled by the red, green and blue arrows of the upper panels,
which are respectively part of the photon ring (3 crossings
of the equatorial plane), lensing ring (2 crossings) and primary
image (1 crossing), in the terminology of~\citet{gralla19}. The half disk filled in black color corresponds to the event horizon.
The black solid half circle of the left panel corresponds to the
Schwarzschild photon
sphere at $r = 3M$. The {black-line delineated white} crescent of the right panel corresponds
to the locus of spherical Kerr orbits for $a=0.8M$, with the
locations of the prograde and retrograde equatorial photon
orbits marked by black dots.
The dashed thick
black line within the crescent corresponds to the spherical orbit
at the inner Boyer-Lindquist radius of the red geodesic.
The thick disk corresponds to the pale red-color region. 
In the Schwarzschild case (left panel), the red
geodesic approaches the photon sphere.
In the Kerr case (right panel), the red geodesic approaches
a spherical orbit at its minimum Boyer-Lindquist radius.
{Both the red and green geodesics belong to the secondary ring.}}
\label{fig:explain_secondary}
\end{figure*}

\begin{figure*}[htbp] 
\centering
\includegraphics[width=0.5\textwidth]{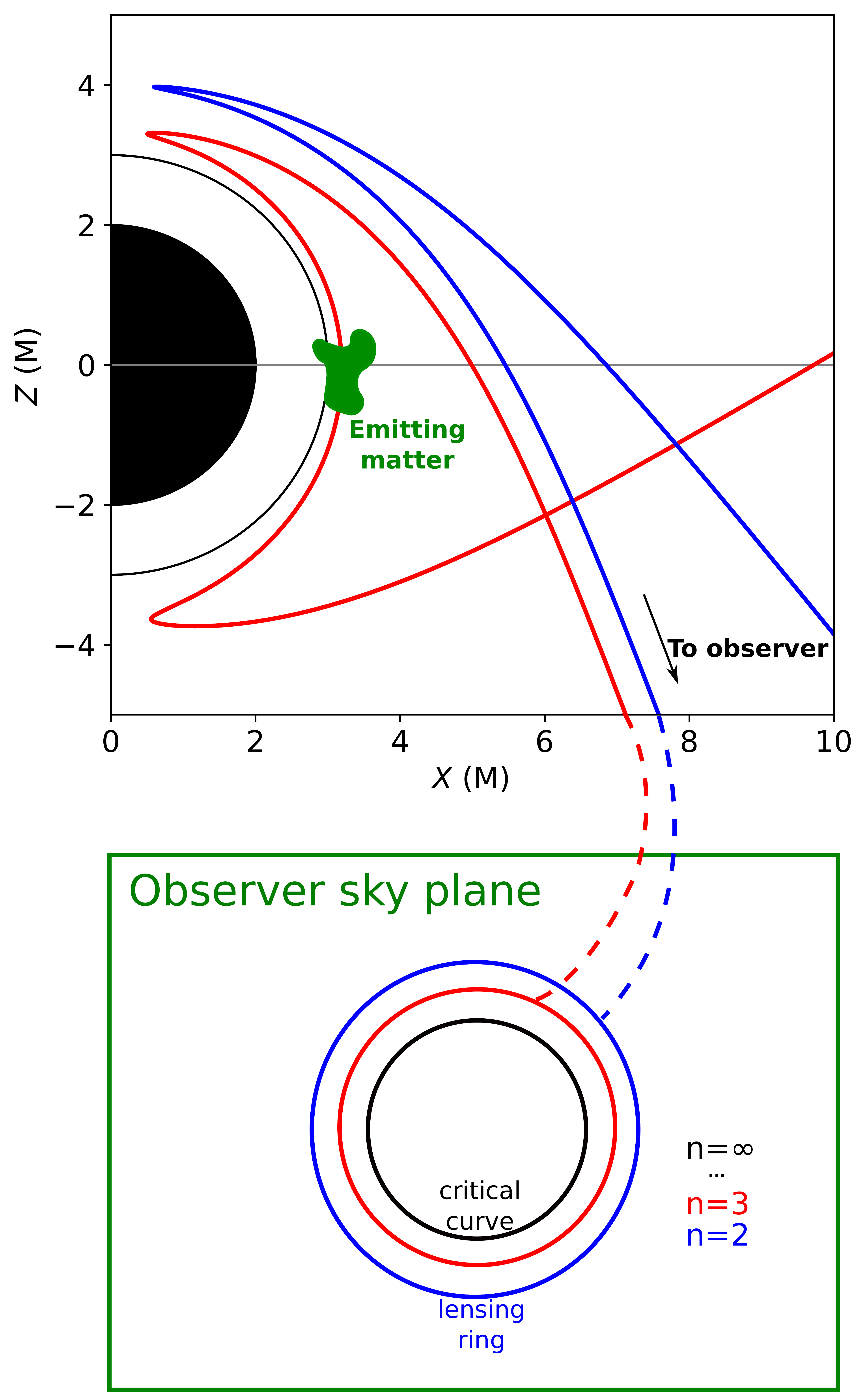}
\caption{{Example of a secondary ring.
    \textit{Top:} Two null geodesics, in Schwarzschild spacetime,
    that are part of the
    $n=2$ (blue) and $n=3$ (red) photon sub-rings (see text),
    $n$ being the number of
    crossings of the equatorial plane. The black disk represents the black hole
    event horizon, and the thin black circle shows the location of the
    unstable spherical photon orbit.
    \textit{Bottom:} The plane of the sky
    of the distant observer. The various concentric rings depict the
    photon sub-rings (see text) corresponding to the different values
    of the number $n$ of crossings of the equatorial plane of the black hole.
    The lensing ring and critical curve (see text) are explicitly labeled.
    The set of sub-rings
    on the observer's sky is not an observable. What is observable is the
    subset of these rings where there is a detectable amount of flux.
    This subset is what we call the secondary ring.
    If only one blob of emitting matter (in green in the top panel)
    is present close
    to the black hole and interesects the red geodesic only, and
    not the blue geodesic, then only the $n=3$ sub-ring will be illuminated
    on sky, and the others will remain dark. The secondary ring
    will then coincide with the $n=3$ subring. This would not be so should
    the blob of emitting matter be situated elsewhere.
    The spacing between the sub-rings on the observer's sky is of course
    very exaggerated and lies within $\lesssim 1~\mu$as for M87*.}
}
\label{fig:subrings_secondaryring}
\end{figure*}

\begin{figure*}[htbp]
\centering
\includegraphics[width=\textwidth]{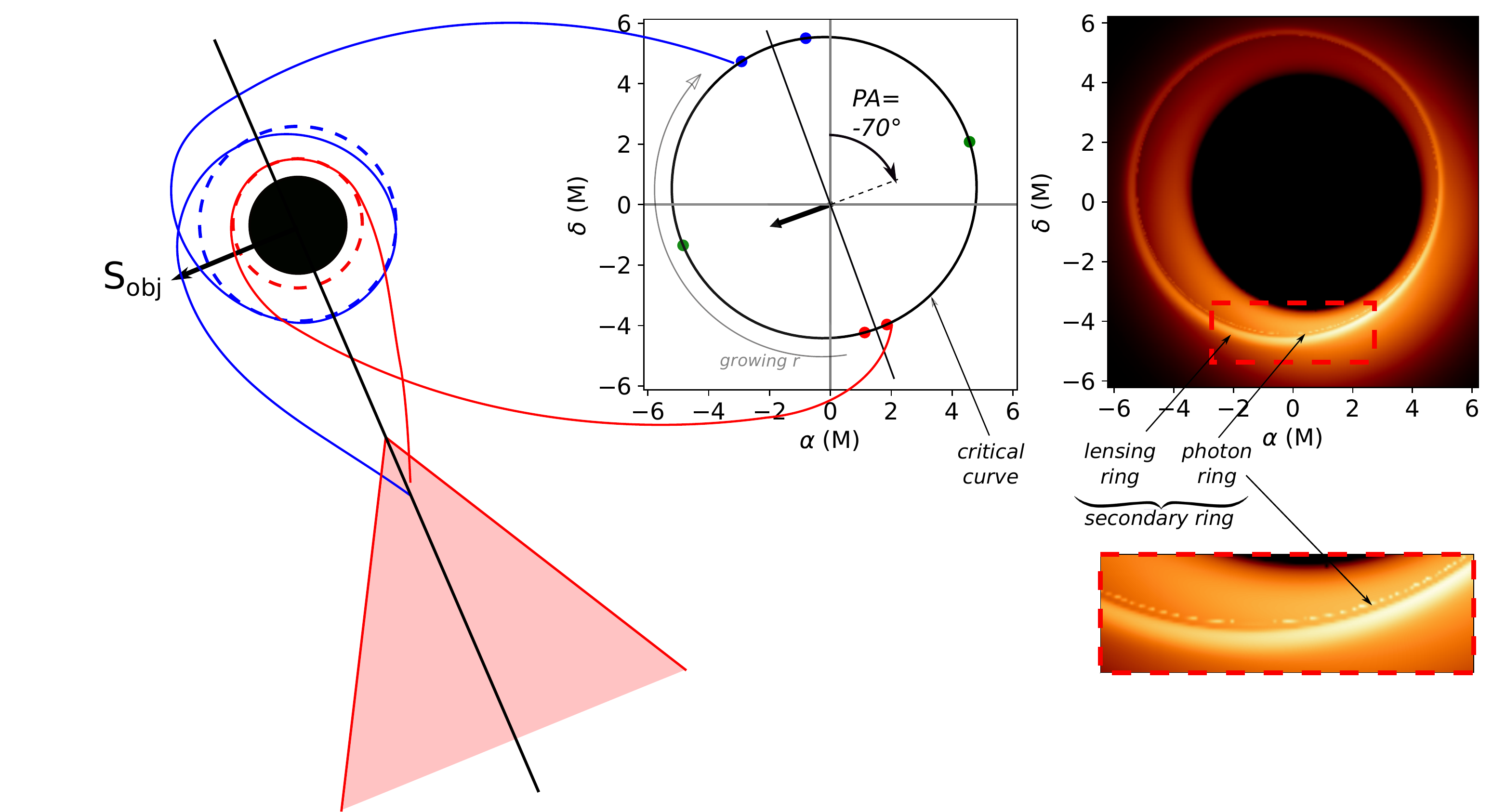}
\caption{The \textit{left drawing} is a sketch of the black hole surroundings,
with the black hole event horizon represented by a black disk,
and the geometrically thick accretion disk represented by the shaded
red region (only one half of the disk is shown to save space). The sketch is rotated in order to have the same orientation
of the black hole spin vector (black arrow) as in the central panel.
The dashed red and blue circle around the black hole are the location
of two Kerr spherical photon orbits (not to scale), 
the radii of which are given below.
The \textit{central panel} is a plot of the critical curve 
on the observer's sky as defined
by the polar equation~\ref{eq:polarcritical},
for a black hole spin of $a=0.8M$ and an inclination angle
of $i=160^\circ$. The critical curve is rotated to account
for the position angle of M87* ($PA=-70^\circ$ as labeled on the panel,
or $290^\circ$ east of north). The projection of the
black hole spin vector is shown by the black arrow (opposite to
the approaching jet projection depicted by the
blue arrow in Fig.~\ref{fig:disk_a0}). Three pairs of points
are shown in red, green and blue along the curve. They are
the image of the Kerr spherical orbits with radii $r=2.31M$ (red),
$2.7M$ (green), and $3.05M$ (blue). The Kerr equatorial
prograde and retrograde photon orbits for $a=0.8M$
are $r_\mathrm{ph,pro}=1.82M$ and $r_\mathrm{ph,retro}=3.81M$.
The gray bent arrow
at the left of the critical curve shows the
direction of increase of the Kerr spherical orbit radius
that the photon is following on its way to the observer (same
evolution on the right side of the critical curve).
Two null geodesics connecting
the red and blue Kerr spherical orbits to the corresponding
points on the critical curve are illustrated between the
left drawing and the central panel. The \textit{right panel}
shows the 
%astrophysical millimeter image
ray-traced image of the model, with the same
field-of-view as in the middle panel. This image is a zoom
on the central $45~\mu$as of the upper-left panel
of Fig.~\ref{fig:disk_a08}. The 4 important notions introduced
in the text: lensing, photon, secondary rings, and critical curve,
are labeled. The zoom on the region of the right panel
surrounded by the dashed red rectangle is shown in the
\textit{lower-right insert}. It allows to better see the difference between
the lensing and photon rings. The critical curve is extremely close in angular
radius to the photon ring (they are impossible to distinguish
with the naked eye when comparing the central and right panels). 
However, the two notions are
different mathematically. In the limit of an infinite resolution,
the photon ring of the right panel would decompose into a sequence of rings, converging to the critical curve.}
\label{fig:illustrate_secondary}
\end{figure*}

In this article, we are interested in the full sequence of
highly-lensed sub-rings on the sky, which incorporates the notions of the
lensing ring, the photon ring, and the critical curve introduced above. However, it is crucial to realize that a pixel
on the observer's camera belonging to one of this sub-rings
will not always contain a detectable amount of flux.
Its flux content, and hence its ability to be considered
as a highly-lensed region on sky, depends on the
corresponding null geodesic interaction with the
accretion flow. As a consequence, the full set of highly-lensed sub-rings
should be seen as a mathematical, theoretical locus
on sky, the flux content of which fully depends
on the accretion flow properties.
We thus introduce the observation-oriented
notion of the \textit{secondary ring} (as opposed
to the primary image) to refer 
to \textit{the region on the observer's sky where the received null geodesics (i) have approached a Kerr spherical photon orbit within $\delta r \lesssim M$ in terms of the radial Boyer-Lindquist coordinate $r$, and (ii) have visited the regions of the accretion flow emitting most of the radiation}.
In this definition, $\delta r$ can be of order $M$ for lensing-ring photons, while $\delta r \ll M$ for photon-ring photons (see the lower panels of Fig.~\ref{fig:explain_secondary}).
{The regions of the flow emitting most of the radiation,
  in our framework, coincide with the inner regions close to
  $r=r_\mathrm{in}$, where all physical quantities are maximal.
  We note that this definition implicitly depends on the orientation
  of the observer with respect to the flow. Indeed, the projection on sky
  of the regions of the flow emitting most of the radiation depends
  on the inclination and position angle.}

Our definition is 
based on more than just the number of crossings of the equatorial plane
by null  geodesics. As discussed in the Appendix~\ref{app:secondary}, a definition based only on the number
of crossings of the equatorial plane is not adequate
as geodesics can cross this plane at very large radii,
and such crossings are not relevant for the definition
of the secondary ring. Moreover, and most importantly,
the secondary ring definition must be linked
to the particular accretion flow model used and its
emission law. This crucial point is illustrated in
Fig.~\ref{fig:explain_secondary}, which shows 
the link between the Kerr spherical
orbits, the accretion flow geometry, 
and the secondary ring of the image.
It first shows that highly-lensed geodesics (with more than 2 crossings of the
equatorial plane close to the black hole) indeed approach a Kerr spherical orbit
in the vicinity of the black hole. Most importantly,
it also shows that not all geodesics that
approach Kerr spherical orbits will correspond to bright pixels
of the image. A secondary-ring geodesic is not only highly bent, but
it is also selected by the fact that it should visit the inner
parts of the accretion flow in order to transport enough flux
(the red geodesic of the lower-right panel of Fig.~\ref{fig:explain_secondary}
is a good example: its spherical-orbit radius is exactly equal
to the radius of the bright inner edge of the disk,
%inner disk radius
allowing to transport a lot of flux).
Consequently, both the polar radius on sky and the azimuthal
flux distribution of the secondary ring are depending
on the properties of the accretion flow; they are not simply
dictated by gravitation.
Should the inner radius
of the disk of the lower-right panel of Fig.~\ref{fig:explain_secondary}
be moved down to the event horizon, the red geodesic would transport a
much smaller amount of flux, and would thus not be
considered as belonging to the secondary ring (the geodesic optical path within the flow would be longer, but this
increase scales as $r$, while the decrease in 
density scales as $r^{2}$, so that the resulting
flux would be smaller).
It has been recently shown by~\citet{gralla19b} that the dependence of the lensing ring polar radius on the accretion flow geometry can reach tens of percent (see their Fig.~5).
We note that this result,
obtained for a geometrically-thin disk, should
be considered as a lower limit in a geometrically-thick disk context.

{Figure~\ref{fig:subrings_secondaryring} illustrates the
notion of secondary ring and highlights its dependence
on the properties of the plasma surrounding the black hole.}

Four important notions have been introduced so far: lensing,
photon and secondary rings, and the critical curve; some of which having non-trivial definitions. Figure~\ref{fig:illustrate_secondary} gives a pedagogical
illustration of these notions. 
We insist on the fact that the only new word that we introduce here, i.e. the notion of secondary ring, is really needed. Indeed, it conveys the crucial idea that highly-lensed features in the image plane are intrinsically depending on the astrophysical accretion model, which does not appear clearly in the definition of other notions (lensing, photon rings).

Let us now discuss more quantitatively our Kerr images.
We note that in Figs.~\ref{fig:disk_a0} and~\ref{fig:disk_a08},
the angular size of the dark central region depends a lot on the
inner radius of the accretion flow. This is in agreement with the 
simple model of~\citet{gralla19}. In particular, the secondary ring
is not the outer boundary of this central dark region when the flow
extends to the horizon.
 On the other hand,~\citet{narayan19}
recently showed that a spherical optically thin flow in a Schwarzschild spacetime
results in a central dark region the angular size of which is
independent of the location of the inner edge of the emitting region.
This discrepancy once again highlights the importance of the careful modeling of the accretion flow for the interpretation of EHT images.

It is also interesting to determine the 
brightness ratio of the secondary ring to the primary image.
We have checked that in the non-rotating case,
the secondary ring weight is of $5\%$
when $r_\mathrm{in} = r_\mathrm{ISCO}$,
$20\%$ when $r_\mathrm{in} = r_\mathrm{H}$ with
azimuthal flow velocity, and
$15\%$ when $r_\mathrm{in} = r_\mathrm{H}$ with
radial flow velocity.
For the $a=0.8M$ case, the secondary ring weight
is of $30\%$ when $r_\mathrm{in} = r_\mathrm{ISCO}$,
$25\%$ when $r_\mathrm{in} = r_\mathrm{H}$ with
azimuthal flow velocity, and
$20\%$ when $r_\mathrm{in} = r_\mathrm{H}$ with
radial flow velocity.
These numbers are obtained following the
methodology presented in Appendix~\ref{app:secondary}.
As explained there, they should be considered as
slightly over-estimated. For comparison, \citet{Johnson2019} characterize the secondary ring to be responsible for $\sim\!10 \%$ of the total flux seen in ray-traced GRMHD simulations, with 
specifically a weight of $20\%$ reported for their Fig.~1. Our results are thus in good agreement
with the more sophisticated GRMHD simulations.

\subsection{Non-axisymetric emission}
\label{sec:nonaxisym}

The flux distribution seen in Fig.~\ref{fig:disk_a0}
and~\ref{fig:disk_a08}, with the south region of the image brighter than the north part, is primarily due to the beaming effect.
This is so because the emission is assumed to be axisymmetric. However, non-axisymmetric
flux distribution is necessarily present in realistic turbulent flows.
It is thus a natural question to ask what should be the condition
on the non-axisymmetry of the emission such that the flux
repartition would be substantially altered.  
To investigate this point, we study a very simple non-axisymmetric feature in our disk model.
We consider that some region of the disk, centered at
a cylindrical radius (defined in Boyer-Lindquist
coordinates by $\rho = r\,\sin\theta$) of $\rho=\rho_0$ and azimuth $\pp=\pp_0$,
with typical extensions $\sigma_\rho$ and $\sigma_\pp$,
will be hotter than the rest by some increment $\Delta T$.
Specifically, we consider that the temperature around
$(\rho_0,\pp_0)$ will read
\be
T(\rho,\pp) = T_\mathrm{axisym}(\rho) + T_0 \, G(\rho,\pp)
\ee
where $T_\mathrm{axisym}(\rho)$ is the axisymmetric temperature
defined in the previous section, $\Delta T = T_0 \,G(\rho,\pp)$,
$T_0$ is a chosen parameter,
and the function $G(\rho,\pp)$ is the following product of Gaussians
\be
G(\rho,\pp) = \frac{1}{2\pi \sigma_\rho \sigma_\pp} \mathrm{e}^{-\frac{1}{2} \left(\frac{\rho-\rho_0}{\sigma_\rho}\right)^2} \, \mathrm{e}^{-\frac{1}{2} \left(\frac{\pp-\pp_0}{\sigma_\pp}\right)^2}.
\ee
To enhance the difference with respect to the axisymmetric case,
we choose $(\rho_0,\pp_0)$ such that this region is located
towards the north on the sky, i.e., opposed to the beamed region.
The parameters $\sigma_\rho$ and $\sigma_\pp$ are chosen such that
the hotter region has a comparable extension on the sky as compared
to the beamed region of the axisymmetric images.

\begin{figure*}[htbp]
\centering
\includegraphics[width=\textwidth]{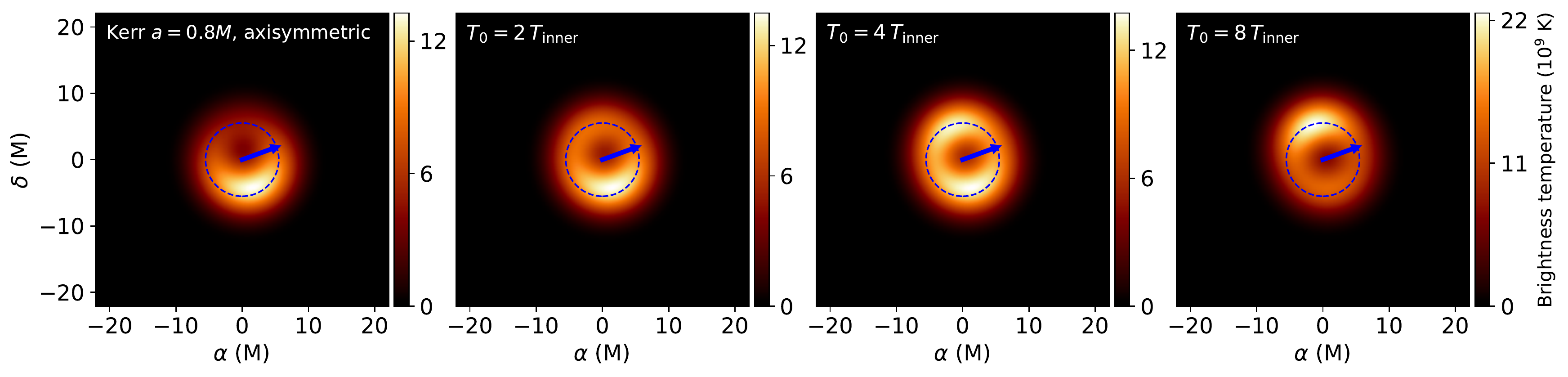}
\caption{Non-axisymmetric disk compared to axisymmetric
  case for spin $a=0.8\,M$. Here, only blurred images are shown.
  The second, third and fourth panel
  from the left are obtained by considering a hotter region
  in the disk defined by a temperature increment of
  $T_0/T_\mathrm{inner} = 2$, $4$, or $8$ respectively.
  This comparison shows that the non-axisymmetry of the flow must
  be substantial (approximately an order of magnitude contrast)
  in order to overcome the beaming effect.
} 
\label{fig:disk_nonaxisym}
\end{figure*}

Fig.~\ref{fig:disk_nonaxisym} shows the images obtained when $T_0$ is varied.
The images indicate that the temperature has to increase by
a factor of around $8$ in order for the non-axisymmetric
structure to overcome the beaming effect. Seeing such an unusually hot coherent component in the GRMHD simulations is rather uncommon. This can be seen in Fig. 9 of \citetalias{EHT5}, where a collection of GRMHD snapshots fitted to the EHT data create a distribution centered around an expected brightness maximum position angle of $\approx 200^\circ$ (about 90$^\circ$ clockwise from position angle of the approaching jet projection) with turbulence related scatter of $\sigma \approx 60^\circ$. Nevertheless, there is a non-zero probability for a very different fitted orientation. It is not entirely clear how accurate are GRMHD models at reproducing the intrinsic turbulence-induced structural variability of a realistic accretion flow in the vicinity of a black hole, as EHT is the first instrument to deliver observational data that could be used to test this.

\section{Emission from a geometrically thick disk in non-Kerr spacetimes}
\label{sec:nonkerr}

In this section, we present
millimeter images of a geometrically thick disk surrounding compact
objects that are different from the standard Kerr black hole. We will first focus on
non-rotating solutions (Minkowski and 
ultracompact star spacetimes, see sections
\ref{sec:minko} and \ref{sec:star}) and
then on rotating solutions (boson star
and Lamy wormhole spacetimes,
see sections \ref{sec:BS} and \ref{sec:lamy}).
Our goal is to determine whether or not the current EHT
data can exclude non-Kerr spacetimes {based on arguments independent of the geometric structure of the accretion flow}. This section presents theoretical images, while the fits to EHT data are discussed
in section~\ref{sec:EHTfit}.

\subsection{Minkowski spacetime}
\label{sec:minko}

We start by considering the most extreme non-Kerr case of a flat
spacetime described by Minkowski geometry. While there may be little physical motivation to consider such an object as a viable alternative to a Kerr black hole, with this exercise we investigate whether \textit{any} spacetime curvature is absolutely necessary to explain the EHT images.
This means that we only consider
the laws of special relativity but discard all general relativistic
effects. This describes what could be thought of as a  Michell-Laplace relativistic black hole~\citep{michell1784,laplace1796}. 
It is a "relativistic" black hole because of
the important addition of special relativity
as compared to the original object. We want to compare
this "flat-spacetime black hole" to a Schwarzschild black hole. 
In both cases,
an accretion disk is assumed to lie in the equatorial plane of
the object, with the same inner radius $r_\mathrm{in} = 6M$. There
is of course no 
{physical motivation}
to terminate the accretion disk at
this radius for our Michell-Laplace relativistic black hole. Our choice is dictated
by the comparison to the Schwarzschild spacetime.
The angular velocity of the emitting matter 
of the Michell-Laplace relativistic black hole is assumed to follow the Newtonian
law $\Omega \propto r^{-3/2}$. 

\begin{figure*}[htbp]
\centering
\includegraphics[width=0.66\textwidth]{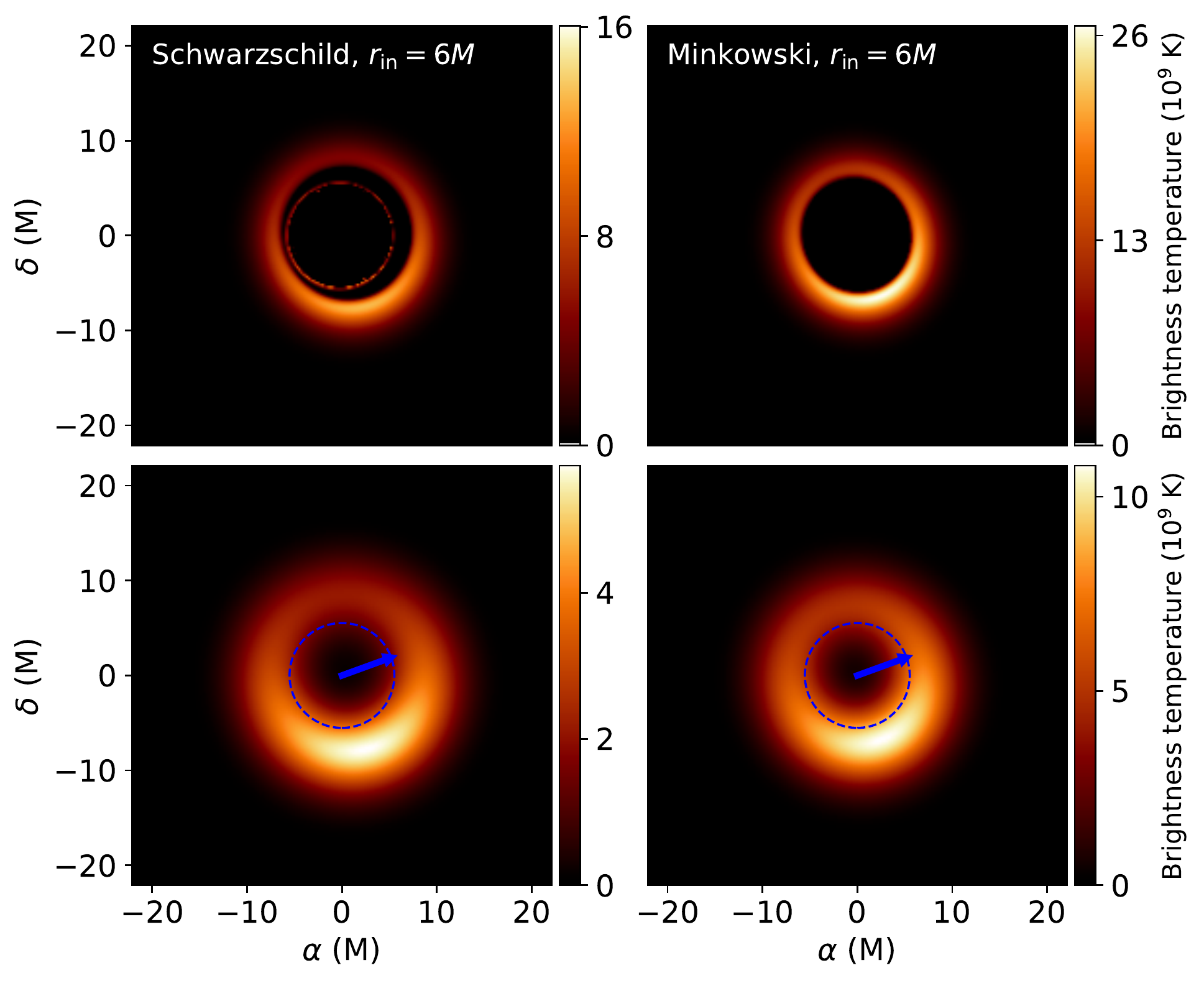}
\caption{Images
  of a geometrically thick accretion disk with inner radius
  $r_\mathrm{in} = 6M$ in a Schwarzschild spacetime (left column)
  or in \textbf{Minkowski} spacetime (right column).
  As in all figures, the bottom row corresponds to the top row images blurred to the EHT resolution ($20\,\mu$as); the dashed blue circle has a diameter
  of $40\,\mu$as (size of the ring feature reported by the EHT) and the blue arrow shows
  the projected direction of the approaching jet.
} 
\label{fig:disk_minko}
\end{figure*}

Figure~\ref{fig:disk_minko} shows a comparison between these two cases.
The high resolution images can be immediately distinguished by the absence of a secondary ring in the Minkowski spacetime. We return to that aspect in section \ref{sec:detect_secondary}, discussing future observational perspectives. Nevertheless, the extreme similarity between the images observed with the EHT resolution (bottom row of Fig. \ref{fig:disk_minko}) is a good illustration that reasoning based exclusively on the image morphology
can tell little about the nature of the central object.

\subsection{Non-rotating ultracompact star}
\label{sec:star}

\begin{figure*}[htbp]
\centering
\includegraphics[width=\textwidth]{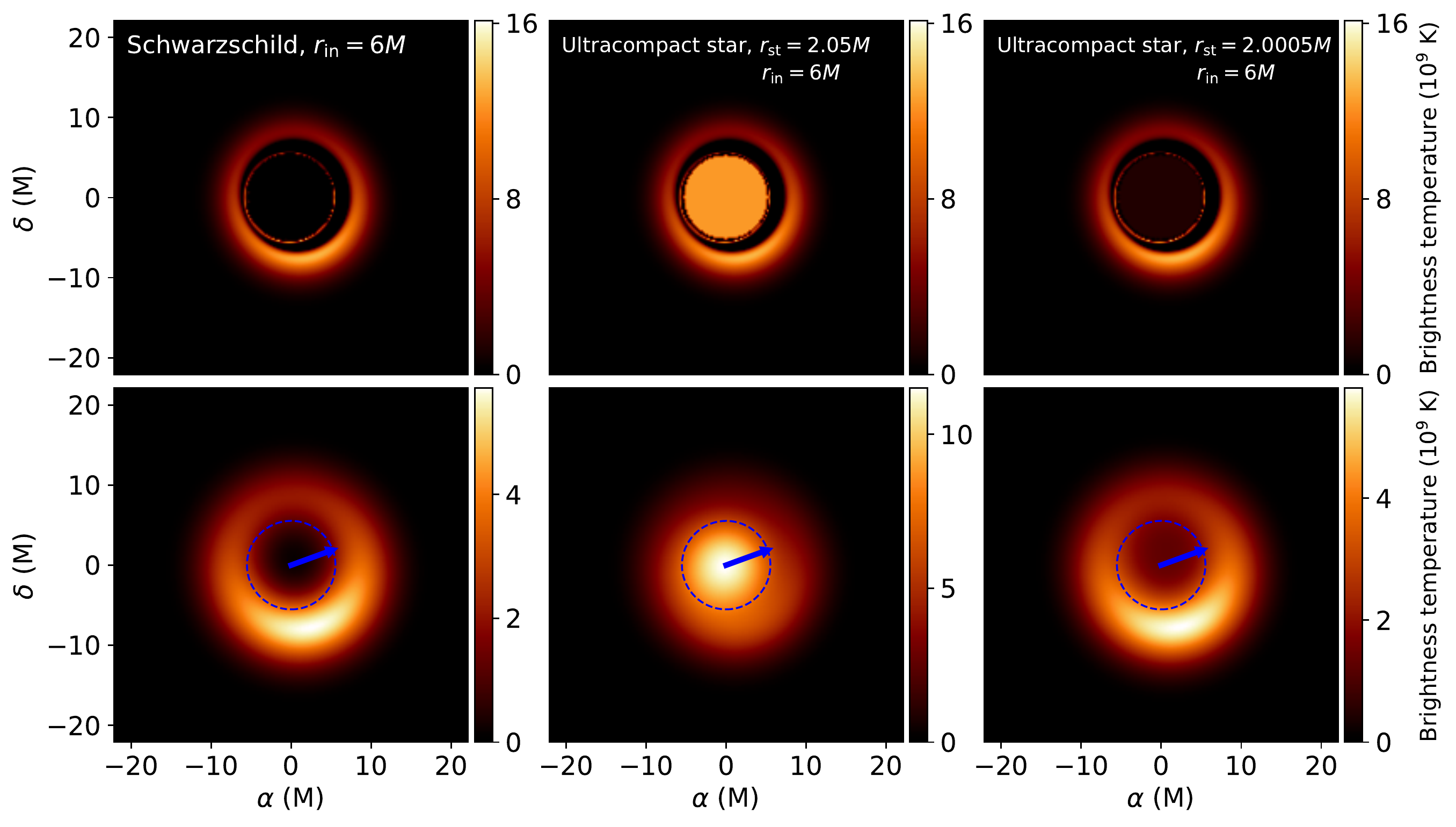}
\caption{Images
  of a geometrically thick accretion disk with inner radius
  $r_\mathrm{in} = 6M$ in a Schwarzschild spacetime (left column),
  in the spacetime of an \textbf{ultracompact star} with surface radius
  $r_\mathrm{st} = 2.05M$ emitting blackbody radiation
  at the inner temperature of the accretion flow
  $T_\mathrm{e,in} = 8 \times 10^{10}$~K (middle column),
  or in the same spacetime as the middle column but with
  $r_\mathrm{st} = 2.0005M$ (right column).
  As in all figures, the bottom row corresponds to the top row images blurred to the EHT resolution of $20\,\mu$as; the dashed blue circle has a diameter
  of $40\,\mu$as (size of the ring feature reported by the EHT) and the blue arrow shows the projected direction of the approaching jet.
} 
\label{fig:disk_CompactStar}
\end{figure*}

We compute here the image
of M87* assuming that the central compact object is not a black
hole but rather an ultracompact non-rotating star with a surface slightly
above the radius of its event horizon. {While we refer to this hypothetical object as a ``star", its only assumed property is the presence of a surface, as we do not consider any internal physics of the object.} Birkhoff's theorem ensures that
the metric at the exterior of this object will be the Schwarzschild
metric, provided the ultracompact star is spherically symmetric.
We note that, should the star rotate,
its exterior metric will not be in general that of Kerr so that the generalization
to a rotating spacetime is not straightforward.

Our ultracompact-star spacetime is defined
as follows.
The star surface is modeled as a spherical
surface of Boyer-Lindquist radial coordinate $r_\mathrm{st} = (2+\epsilon)M$
with $\epsilon \ll 1$ in a Schwarzschild spacetime.
The star's surface is assumed to be fully optically thick
so that its interior (which is not properly
modeled in our setup) is never visited by any photon.

The star's surface is assumed to emit blackbody radiation
at the temperature of the inner accretion flow,
$T_\mathrm{e,in}$, here assumed to be $T_\mathrm{e,in,0} = 8\times 10^{10}$~K. This is of course
a very strong assumption. It is likely, however, that this
surface should be thermalized given that null
geodesics 
are highly curved when emitted at the star's surface and
thus efficiently couple different parts of the surface~\citep{broderick09}.
Moreover, 
{the considerations presented here} will not be qualitatively affected if the
surface temperature is not exactly equal to the inner accretion
flow temperature.

Let us now discuss more quantitatively the radiation
emitted at the star's surface. The observing frequency $\nu_\mathrm{obs}$
is fixed in the whole article to the EHT observing
frequency, $\nu_\mathrm{obs,0} = 230$~GHz.
The emitted frequency at the star's surface is simply
related to that by $\nu_\mathrm{em} = \nu_\mathrm{obs}/g$,
where $g = (1 - 2M/r_\mathrm{st})^{1/2}$ is the redshift factor,
which decreases to $0$ as $r_\mathrm{st}$ approaches the
Schwarzschild event horizon.
The Planck function $B_\nu(\nu,T_\mathrm{e,in})$ peaks at a very high
frequency of $\nu_\mathrm{max} \approx 5 \times 10^{21}$~Hz.
The emitted frequency reaches this value for $(r_\mathrm{st} - 2M)/M = \epsilon \approx 10^{-17}$.
In the following we will thus safely assume that
the Planck function is in its Rayleigh-Jeans regime.
Using the frame-invariance of $I_\nu / \nu^3$,
we can thus express the observed specific intensity as
\bea
I_\nu^\mathrm{obs} &\approx& \frac{2 \nu_\mathrm{obs}^2}{c^2}\,k T_\mathrm{e,in} \left(1 - \frac{2M}{r_\mathrm{st}} \right)^{1/2} \\ \nn
&\approx& \frac{2 \nu_\mathrm{obs}^2}{c^2}\,k T_\mathrm{e,in} \sqrt{\frac{\epsilon}{2}} \ , \\ \nn
\eea
where $k$ is the Boltzmann constant, $c$ is kept explicitly
for clarity, and we have used the assumption that $\epsilon \ll 1$.
We want this observed specific intensity, corresponding to the interior of the secondary ring in the ray-traced images of Fig.~\ref{fig:disk_CompactStar}, to be equal
to some fraction $1/\kappa$ of the maximum observed
specific intensity from the accretion disk, 
$I_\nu^\mathrm{max}$. We thus write
\be
\frac{2 \nu_\mathrm{obs}^2}{c^2}\,k T_\mathrm{e,in} \sqrt{\frac{\epsilon}{2}} = \frac{1}{\kappa} I_\nu^\mathrm{max}
\ee
and
\be
\epsilon = \frac{c^4 \left( I_\nu^\mathrm{max}\right)^2}{2 k^2 \nu_\mathrm{obs}^4 T_\mathrm{e,in}^2 \kappa^2}.
\ee
When considering a Schwarzschild black hole surrounded
by a thick disk with $r_\mathrm{in} = 6M$ (see Fig.~\ref{fig:disk_a0},
top-left panel), the maximum observed specific intensity
from the accretion disk is of the order of $I_{\nu,0}^\mathrm{max} \approx 2\times 10^{19}$ Jy\,$\cdot$\,srad$^{-1}$. Fixing $\kappa = 10$ (the stellar
surface emission is negligible), corresponding to the dynamic range of the EHT images \citepalias{EHT4}, we derive $\epsilon = 0.0005$, and for $\kappa = 1$ (the stellar surface dominates) we find $\epsilon=0.05$.
The following equation gives a practical expression for $\epsilon$
\be
\epsilon \le \frac{0.05}{\kappa^2}\left( \frac{\nu_\mathrm{obs}}{\nu_\mathrm{obs,0}} \right)^{-4} \left(\frac{T_\mathrm{e,in}}{T_\mathrm{e,in,0}} \right)^{-2} \left(\frac{I_\nu^\mathrm{max}}{I_{\nu,0}^\mathrm{max}}  \right)^2 \,
\label{eq:epsilon}
\ee
{that can be understood as a joint constraint on $\epsilon$ and the surface temperature.} 
Figure~\ref{fig:disk_CompactStar} shows the image of an accretion disk with $r_\mathrm{in} = 6M$ surrounding a Schwarzschild black hole (left panel), and an ultracompact star with
surface radius defined by $\epsilon = 0.05$ (middle panel) or $\epsilon = 0.0005$ (right panel). Provided that $\epsilon$ is small enough,
there is no noticeable difference between the Schwarzschild and
ultracompact-star cases. We bring up future perspectives of constraining $\epsilon$ in section \ref{sec:constraining-epsilon}.

Although we do not discuss gravastars~\citep{mazur04} in this article,
we note that non-rotating gravastars would lead to similar images as our Fig.~\ref{fig:disk_CompactStar},
because in both cases a near-horizon surface is present and the external spacetime is Schwarzschild.

\subsection{Rotating boson star}
\label{sec:BS}

In this section, we consider the spacetime of a rotating
boson star, as computed by~\citet{grandclement14}. We are
modeling what is known as a mini boson star, in the
sense that we do not consider any self-interaction between the bosons.
Boson stars
are composed of an assembly of spin-0 bosons consituting a
macroscopic quantum body that evades collapse to a black
hole by means of Heisenberg uncertainty relation~\citep{liebling17}.
Boson stars have no hard surface, no event horizon, and
no central singularity. As such they are extremely different
from black holes and are a good testbed for horizonless spacetimes~\citep{vincent16}.

{A boson star is defined by two parameters, $k \in \mathbb{N}$
and $0 \leq \omega \leq 1$~\citep[see][for details; note that $\omega$ is in units of $m_b c^2 / \hbar$ where $m_b$ is the mass of the boson]{grandclement14}.
The angular momentum of a boson star is quantized and proportional to the integer $k$ because
of the quantum nature of the object. The parameter $\omega$ is related to the compactness of
the star, with compactness increasing when $\omega$
approaches $0$. Here, we consider a boson star
defined by $(k=1, \omega=0.77)$, which has already been
discussed in~\citet{vincent16}. For $k=1$ boson stars
may or may not have  photon orbits depending on the
value of $\omega$~\citep{grandclement17}. If  photon orbit exist, there must
exist at least two of them, one of them being stable, leading to a~questionable stability of the spacetime~\citep[][this statement is actually much more general and applies to any axisymmetric, stationary solution of the Einstein field equations with a matter content obeying the null energy condition]{cunha17b}. The $(k=1, \omega=0.77)$ boson star spacetime is interesting because there is no known reason to question its stability. In particular, it has niether a stable photon orbit, nor an ergoregion. {Its parameters translate to a spin of $a=0.8M$}, the same as the Kerr black hole discussed in section~\ref{sec:kerr}.}

\begin{figure*}[htbp]
\centering
\includegraphics[width=\textwidth]{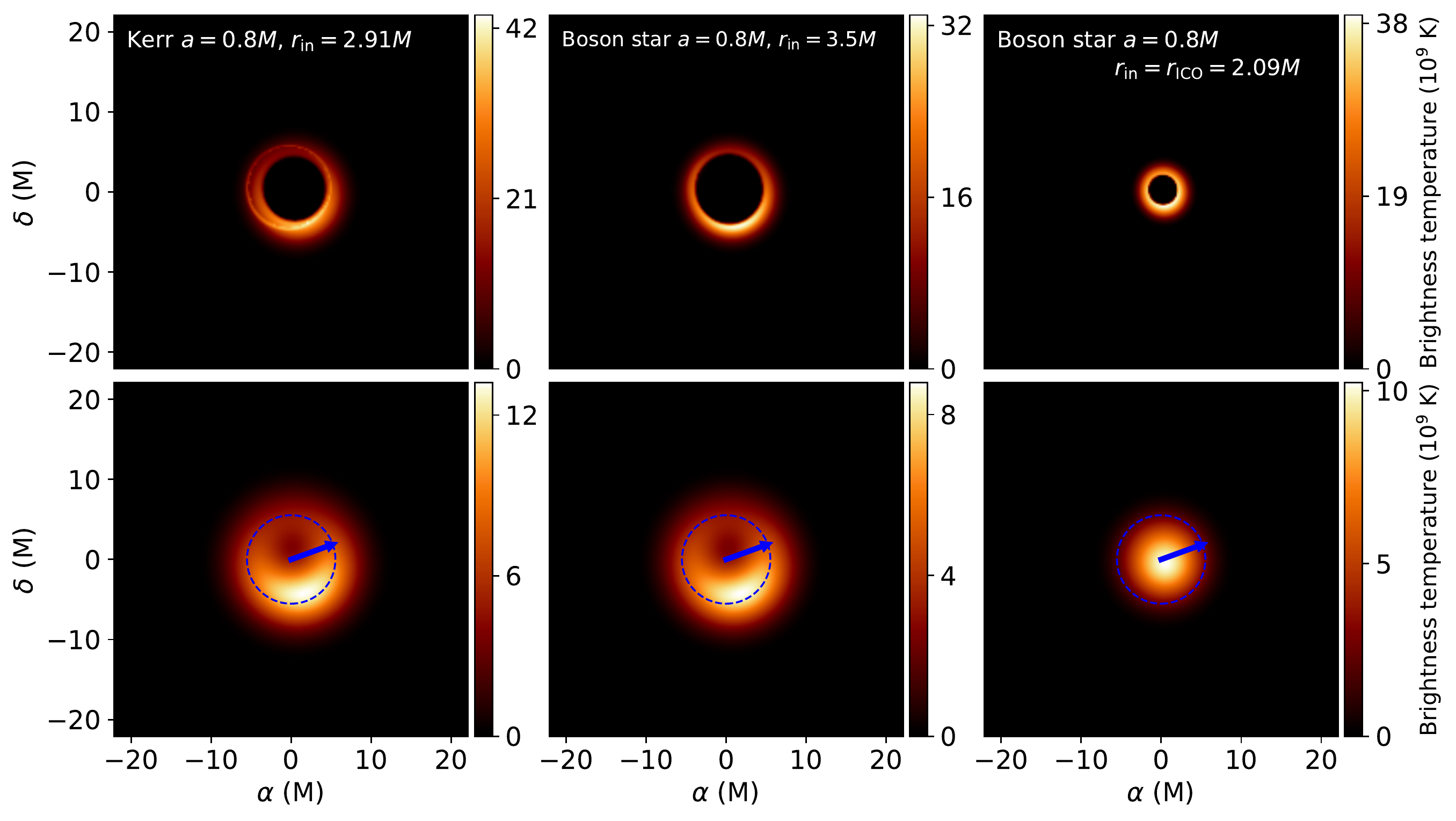}
\caption{Images
  of a geometrically thick accretion disk with inner radius
  $r_\mathrm{in} = 2.91M$ in a Kerr spacetime with
  spin parameter $a=0.8M$ (left column). 
  Same image, but with a disk inner radius at $r_\mathrm{in} = 3.5M$, in a rotating \textbf{boson star} spacetime defined
  by $(k=1,\omega=0.77)$, which corresponds to the
  same value of the spin parameter (middle column). Same boson-star spacetime with $r_\mathrm{in} = 2.09M$, corresponding to the
  innermost circular orbit of that spacetime (right column).
  As in all figures, the bottom row corresponds to the top row images blurred to the EHT resolution of $20\,\mu$as; the dashed blue circle has a diameter
  of $40\,\mu$as (size of the ring feature reported by the EHT) and the blue arrow shows
  the projected direction of the approaching jet.
} 
\label{fig:disk_BS}
\end{figure*}

Figure~\ref{fig:disk_BS} shows a comparison between the
image of a geometrically thick accretion disk surrounding a Kerr
black hole and the rotating boson star discussed above.
For the Kerr spacetime, the inner radius of the accretion disk is fixed at the ISCO, $r_\mathrm{in,Kerr} = 2.91M$.
For the boson star spacetime, using the same inner radius leads
to a slightly too small image on sky. We thus increased it
to $r_\mathrm{in,BS} = 3.5M$ in order to match as closely
as possible the Kerr image. The inner number density
is chosen accordingly, following our $r^{-2}$ power law.
Choosing a different inner radius for the two
spacetimes is not an issue, our goal being only to determine whether
a boson-star spacetime can mimick a Kerr spacetime.
The emitting matter of the boson star spacetime
is following circular timelike geodesics
of the boson-star metric,
the equation of which can be found in~\citet{grandclement14}.
These authors have analyzed the stability of timelike circular geodesics for boson stars. They show that all
circular timelike geodesics are stable for boson stars, so that
it is sufficient to speak of the innermost circular
orbit (ICO).
Our $(k=1,\omega=0.77)$ boson star has an ICO at
$r_\mathrm{ICO} = 2.09M$. Our choice of $r_\mathrm{in,BS} = 3.5M$
means that the inner disk radius is at $\approx 1.7$ times
the ICO radius for the boson star spacetime.
For comparison, we also show the image corresponding
to a choice of $r_\mathrm{in,BS} = r_\mathrm{ICO}$
in the right column of Fig.~\ref{fig:disk_BS}.

The boson-star case with a larger inner
radius of the accretion flow leads to a blurred image very
similar to Kerr, given that the thin secondary ring of the Kerr image is
washed out by the limited resolution of the observations. On the other hand, setting the inner radius at the ICO leads to a much smaller image on
sky (assuming the same mass), which results in a blurred image
very different from Kerr. {This shows that it is the accretion flow properties that matter when comparing a boson star to a Kerr black hole. Modifying the spacetime geometry alone is not sufficient in order to 
produce an observationally different image independently on the accretion flow geometry.
This demonstrates that more sophisticated simulations, connecting general relativity and the accretion flow magnetohydrodynamics, may be necessary to convincingly discuss the observable differences between black holes and other compact objects.}

Recently,~\citet{olivares2018} published the first GRMHD simulation
of an accretion flow surrounding a non-rotating boson star.{{
They computed the associated $230$~GHz image, taking into
account physical parameters typical of the Sgr~A* environment, concluding that it is possible to distinguish a boson star
from a black hole by comparing the non-rotating boson-star
image to {a Schwarzschild and $a=0.937M$ Kerr images}. They reported the boson-star image
to be more compact and symmetric, similarly as the results we present in the last column of Fig. \ref{fig:disk_BS}, as a consequence of a gas accumulation at small radii. However, this picture may be different for a fast-spinning boson star {(notice, that there are no slow-rotating boson stars)}. Answering this question requires 
further GRMHD studies, that might in particular be able to discuss the jet power delivered by a boson star.}}

\subsection{Lamy {spinning} wormhole}
\label{sec:lamy}

In this section we consider the 
rotating wormhole solution first described
in~\citet{lamy18}, that we will hereafter refer
to as Lamy wormhole. This solution was found by generalizing the spherically-symmetric regular (i.e. singularity-free) black hole
solution of~\citet{hayward06} to the
rotating case. This metric takes the
same form as the Kerr metric expressed in
Boyer-Lindquist coordinates, but with the constant
$M$ replaced by the function
\be
M(r) = M \frac{|r|^3}{|r|^3 + 2Mb^2}
\ee
where $b$ is a charge homogeneous to
a length (it is expressed in $M$ units with our conventions). In the original Hayward metric,
$b$ is interpreted as a scale at which quantum gravity effects would act and regularize the
classical singularity. It should therefore
typically take extremely small values. 
However, this parameter has been reinterpreted
by~\citet{fan16} as the magnetic charge associated to a magnetic monopole in a
nonlinear electrodynamics theory that
sources the Hayward metric. In this
context, $b$ can take macroscopic values.

Here, we consider only one pair of values 
for the spin parameter and charge,
$a=0.8M$ and $b=M$. This choice fully 
specifies the metric. It can be shown that
this spacetime corresponds to a rotating
wormhole~\citep{lamy18}. In particular,
it has no event horizon, and of course
no curvature singularity. The topology of this
spacetime corresponds to two asymptotically
flat regions, one with $r>0$ and one
with $r<0$, connected by a throat 
at $r=0$. The energy conditions are violated
in the full region $r<0$,
however the stress-energy tensor decreases
fast to zero when $|r|$ increases so that
the exotic matter is concentrated near the throat. This spacetime is
very exotic. It is, however, quickly converging to Kerr away from
$r=0$ (typically, for $r \gtrsim 10M$, the metric is Kerr; the relative difference of $g_{tt}$
for instance is less than $0.05\%$ in the
equatorial plane for $r > 10M$). Thus, a Lamy wormhole can be seen
as an interesting testbed for the
wormhole-like non-Kerrness of spacetime.
The final important property of this spacetime
(as well as all Lamy spacetimes) is that they
admit spherical photon orbits
similar to Kerr's. Their locus depends of course
on the values of $a$ and $b$.
They are analyzed in~\citet{lamy18b}.

\begin{figure*}[htbp]
\centering
\includegraphics[width=0.66\textwidth]{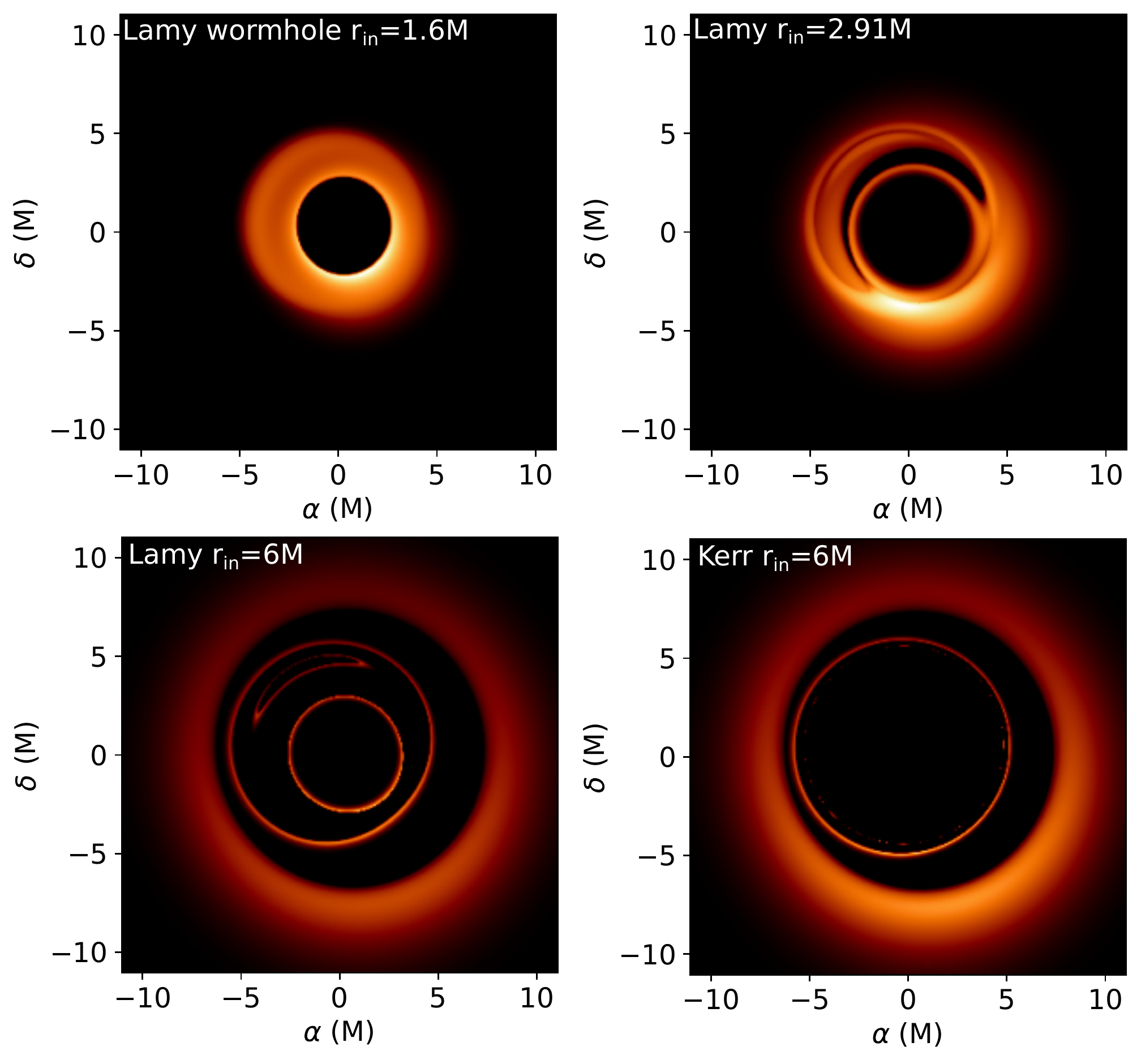}
\caption{The two upper panels and the lower-left panel
show three images with a field of view of $80\, \mu$as of a thick disk surrounding
a \textbf{Lamy wormhole} with spin $a=0.8M$
and charge $b=M$. The inner disk radius is
at $r_\mathrm{in} = 1.6M$ (ISCO radius of
the Lamy spacetime, {upper-left panel}),
$2.91M$ (ISCO radius of the Kerr spacetime
with spin $a=0.8M$, upper-right panel), or
$6M$ (lower-left panel). The lower-right panel shows the image of a thick disk 
with inner 
radius $r_\mathrm{in} = 6M$,
computed with the same field of view, surrounding
a Kerr black hole with spin $a=0.8M$.
In these panels, the image resolution is $300 \times 300$ pixels.
} 
\label{fig:lamy_theo}
\end{figure*}

Figure~\ref{fig:lamy_theo} shows three
images of a thick disk surrounding our
Lamy wormhole, compared to a Kerr image.
The lower-left panel is interesting in order to understand the highly-lensed central part of the image, as it is not confused with the primary image.
The striking feature of the highly-lensed part
of this panel is the existence of two rings, 
and of a crescent in between the rings.
These features are also noticeable in the 
upper-right panel, although less clear as they
overlap with the primary image.
These features are due to extreme light
bending in the central regions of the
Lamy spacetime, due to the existence
of spherical photon orbits. They are absent
in the boson-star image in Fig.~\ref{fig:disk_BS}, as the $k=1$, $\omega=0.77$ boson star has no photon orbits. 

\begin{figure}[htbp]
\centering
\includegraphics[width=0.4\textwidth]{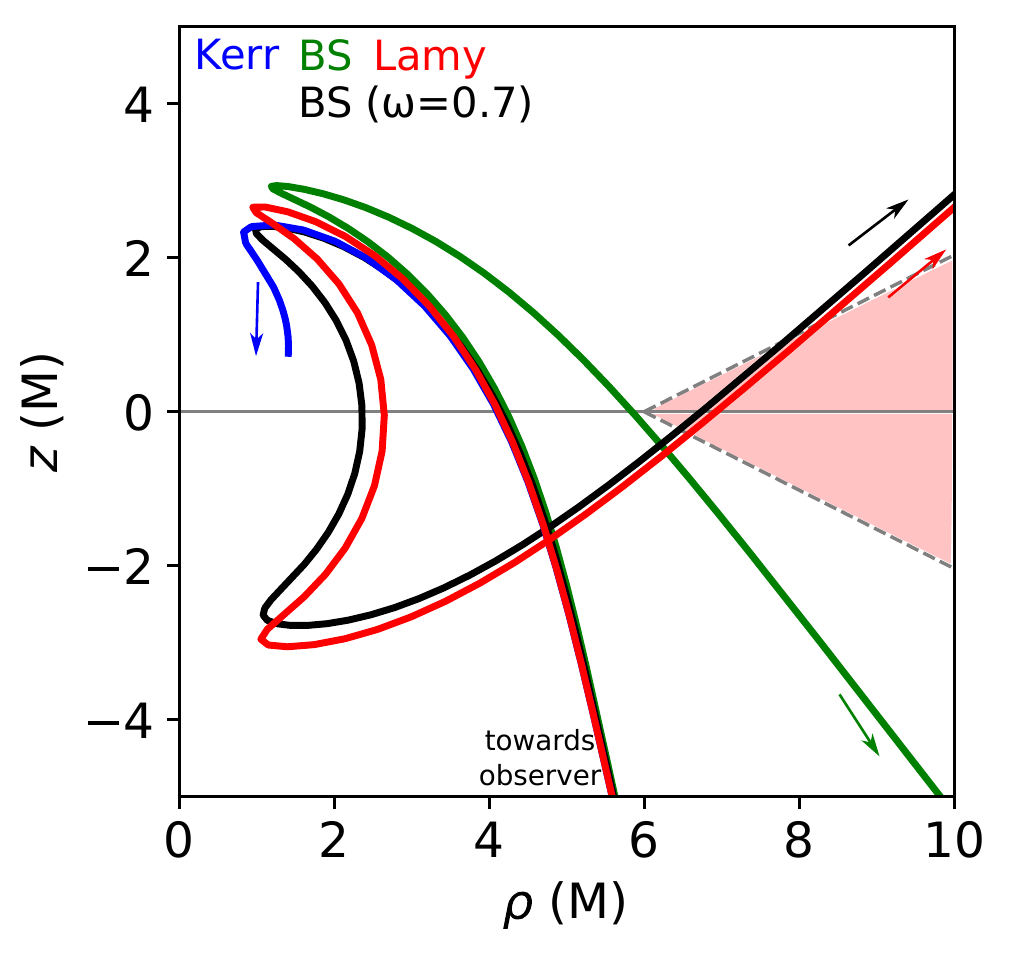}
\caption{Four geodesics corresponding to the same screen pixel for a Kerr (blue; integration stopped
when approaching the event horizon), boson-star (green for the $k=1$,
$\omega=0.77$ boson star considered in this
paper; black for a more compact $k=1$,
$\omega=0.7$ boson star), or Lamy (red) spacetime. The selected pixel is one of
that forming the crescent-shape highly lensed
feature visible on the lower-left panel of Fig.~\ref{fig:lamy_theo}. The arrows represent the direction of backward-in-time ray-tracing integration. The observer is thus at a large negative value of $z$. The disk inner radius in all cases is at $r_\mathrm{in} = 6M$.} 
\label{fig:geod_compare}
\end{figure}

In order to go one step further in the analysis 
of the impact of photon orbits on the image,
we have considered a more compact boson star 
spacetime, with $k=1$ and $\omega=0.70$,
which has been already studied in~\citet{vincent16}. This spacetime has
photon spherical orbits.
Figure~\ref{fig:geod_compare} shows null
geodesics corresponding to one of the
bright pixels of the {inner} crescent feature of the
lower-left panel of Fig.~\ref{fig:lamy_theo}
computed in Lamy, Kerr, and the two different
boson star spacetimes. It highlights the
close similarity of the geodesics corresponding
to the two horizonless spacetimes with photon orbits (Lamy in red and boson star with $k=1$, $\omega=0.70$ in black). Both of them lead
to a very big change of the Boyer-Lindquist
$\theta$ coordinate of the null geodesic
before and after approaching the compact object.
On the contrary, 
the horizonless spacetime with no photon orbit
(boson star with $k=1$, $\omega=0.77$, in green) 
leads to a very different geodesic with
much smaller change of $\theta$ when
approaching the compact object. 
Appendix~\ref{app:BSLamy} shows that the similarity
between the boson star ($k=1$, $\omega=0.70$) and Lamy spacetimes is not restricted to the
particular geodesic represented in Fig.~\ref{fig:geod_compare}. The complete
images are extremely similar and possess
a comparable {inner} crescent feature (see Appendix \ref{app:BSLamy}).% \mw{cite appendix here?}
Note that such a crescent feature was also noticed
for edge-on views in these two spacetimes by~\citet{vincent16} and~\citet{lamy18}.
It is thus plausible that such features are
characteristic of a large class of horizonless spacetimes
with photon orbits.

\begin{figure*}[htbp]
\centering
\includegraphics[width=\textwidth]{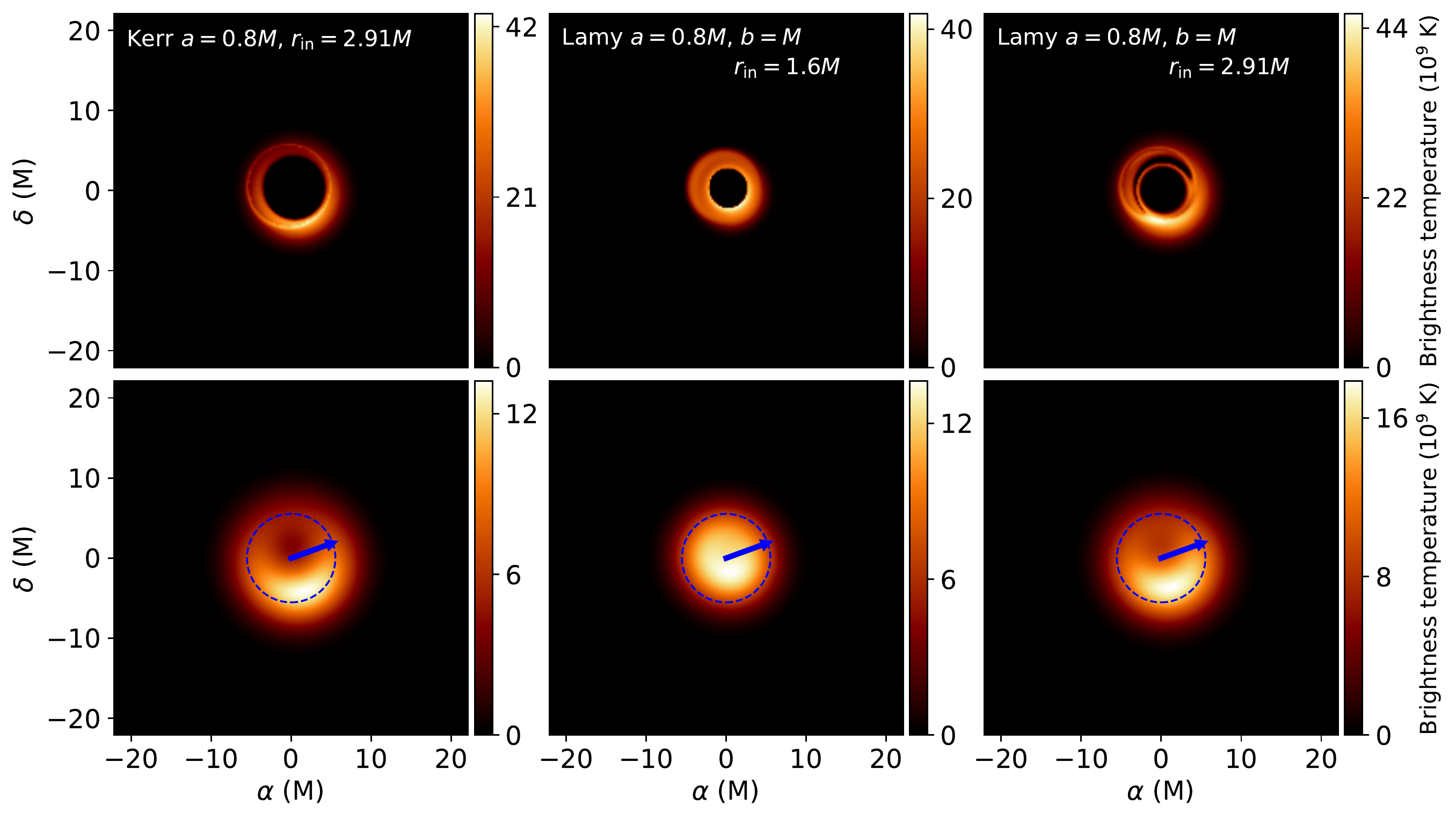}
\caption{Images
  of a geometrically thick accretion disk with inner radius
  $r_\mathrm{in} = 2.91M$ in a Kerr spacetime
  with spin $a=0.8M$ (left column), or in a \textbf{Lamy wormhole} spacetime with spin $a=0.8M$ and charge $b=M$.
  The inner disk radius is at 
  $r_\mathrm{in} = 2.91M$ for the left and right panels, and at $r_\mathrm{in} = 1.6M$ for the central panel.
  As in all figures, the bottom row corresponds to the top row images blurred to the EHT resolution of $20\,\mu$as; the dashed blue circle has a diameter
  of $40\,\mu$as (size of the ring feature reported by the EHT) and the blue arrow shows
  the projected direction of the approaching jet.} 
\label{fig:disk_LamyKerr}
\end{figure*}

Figure~\ref{fig:disk_LamyKerr} compares
the EHT-like images obtained for
Kerr and Lamy spacetimes.
It shows that the complex features of the
Lamy spacetime are partially lost
when blurred at the EHT resolution.
Still, there is a clear excess of flux in
the central fainter region as compared
to Kerr. Given that the non-Kerrness of
Lamy spacetime depends directly on the charge $b$,
it would be possible to derive a constraint
on this parameter by performing fits of various Lamy spacetimes with different
values of the charge. Such a constraint goes
beyond our current analysis.

\section{Fitting models to the EHT data}
\label{sec:EHTfit}

\begin{figure*}[htbp]
\centering
\includegraphics[width=0.99\textwidth]{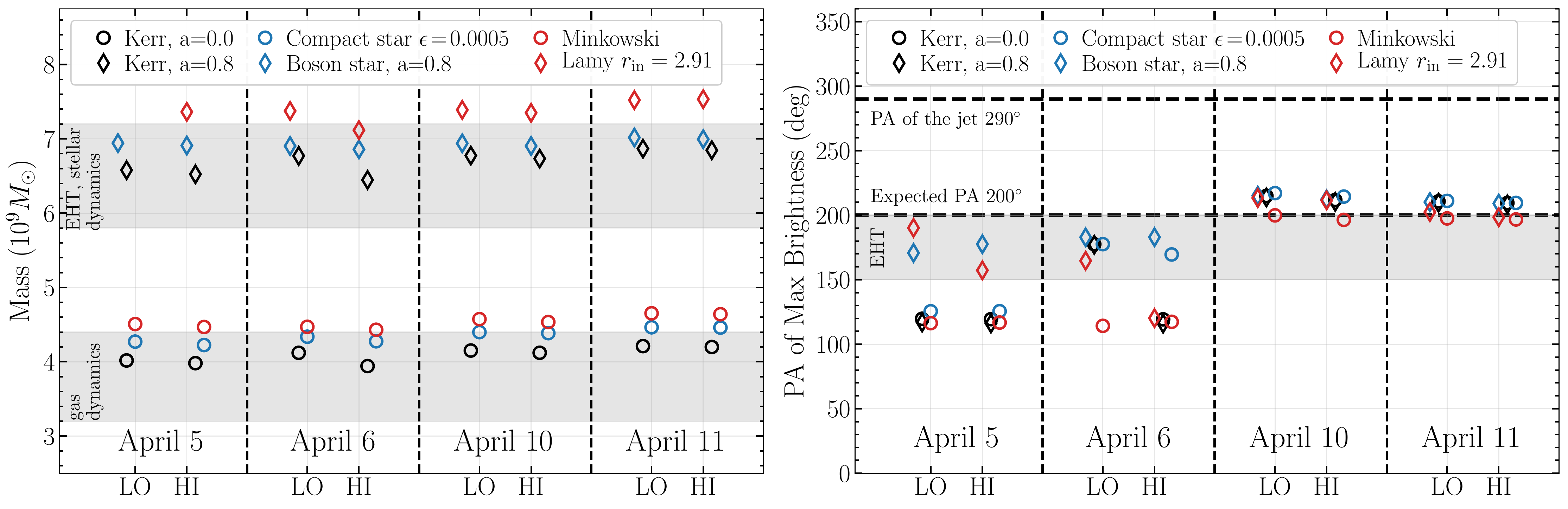}
\caption{Results of fitting the models of section~\ref{sec:nonkerr} to the EHT data sets. Gray bands denote previous measurements of mass and orientation of the M87*. \textit{Left:} Masses of the best-fit models. Different models with spin $a=0.8$ give mass measurement consistent with the one reported by the EHT. Models with zero spin give systematically inconsistent mass estimate. \textit{Right:} Position angles (east of north) of the brightest region in the best-fit models. All models constrain the brightness maximum to be located in the south of the source image, consistently with the EHT results. 
} 
\label{fig:data_fits}
\end{figure*}

{Up to this point we have only discussed the differences between Kerr
and non-Kerr images based on qualitative image-domain comparison. It is important to notice that we did not consider the sparsity-related limitations of the EHT image reconstruction capabilities \citepalias{EHT4}. Effectively, our images represented the actual view of the model at the assumed resolution, without any reconstruction-related distortions.
In contrast, this section is devoted
to comparing models of different compact objects directly to the M87* observational Fourier domain data.}

\begin{figure}[htbp]
\centering
\includegraphics[width=0.99\columnwidth]{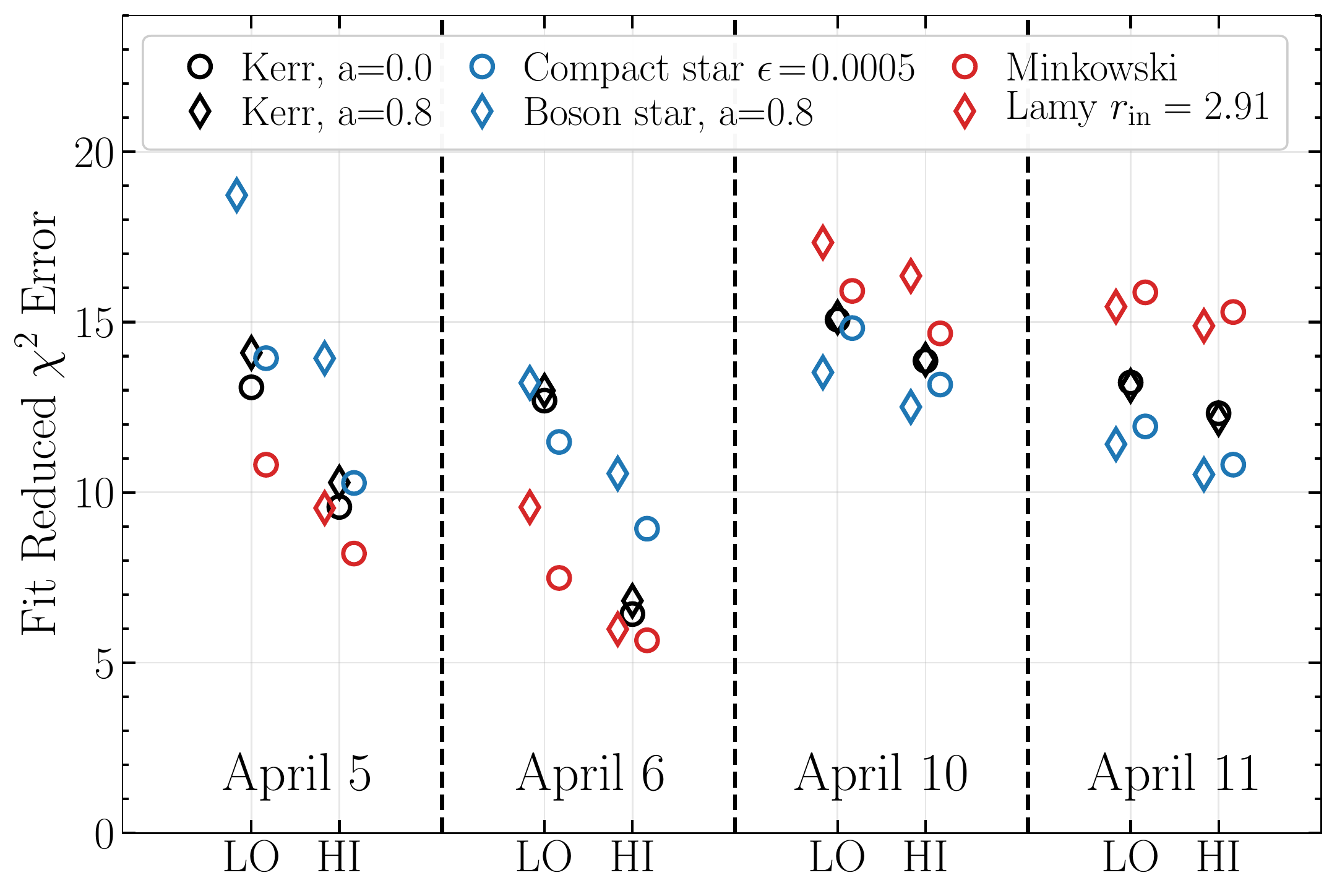}
\caption{Quality of the fits measured by the reduced $\chi^2$ errors resulting from fitting to the interferometric closure quantities in the EHT data sets.
} 
\label{fig:data_fits_quality}
\end{figure}

The total intensity data from 2017 EHT observations of M87* have been publicly released\footnote{\url{https://eventhorizontelescope.org/for-astronomers/data}}. The data consist of 4 independent days of observations in 2 independently recorded and processed frequency bands \citepalias[HI and LO,][]{EHT3}. We performed fitting of the models presented in sections \ref{sec:kerr}-\ref{sec:nonkerr} to all 8 released EHT datasets. As is the case in very long baseline interferometry (VLBI), data correspond to the sparsely sampled Fourier transform of the images on the sky \citep{TMS}. Because of the sparsity limitations, sophisticated postprocessing is required in order to reconstruct the corresponding image \citepalias{EHT4}. While Fourier domain (referred to as visibility domain in this context) data offer well understood error budget, reconstructed images may suffer from the difficulty to assess systematic uncertainties. This is why all quantitative model fitting should take place in the visibility domain. In our case we are sampling the ideal (unblurred) model images using a synthetic model of the EHT array, utilizing the exact coverage of the M87* observations in the 2017 EHT campaign, and the expected magnitude of uncertainties. That part of the work was performed in the framework of the \texttt{eht-imaging} library \citep[][\url{http://github.com/achael/eht-imaging}]{Chael2016,eht-imaging}.
%\citep[see][for an example how not to perform such analysis]{Yosuke2018}}. 
For the crude fitting procedure that we utilize in this paper, we consider scaling, rotation, and blurring of the model images, minimizing the reduced $\chi^2$ errors calculated against robust interferometric closure quantities \citep[closure phases and log closure amplitudes,][]{ closures}. In each iteration of the error minimization procedure, the updated model is sampled and compared with the observed closure data. The procedure of selecting the linearly independent set of closure data products and defining the exact form of the minimized error functions follows that described in \citetalias{EHT4} and \citetalias{EHT6}. From the estimated scaling parameter, we recover the mass $M$ of the model best fitting the data. The rotation parameter allows to calculate the position angle of the bright feature found in the best-fitting model. Note that both scale and orientation of the image were fixed in the discussions in previous sections to the values given in Tab. \ref{tab:properties}. The results of the fitting procedure are summarized in Fig. \ref{fig:data_fits}. All models individually give very consistent mass estimates for all days and bands, indicating that the errors are not of random statistical character, but rather are dominated by the systematic model uncertainties.
Zero-spin models consistently result in much lower estimated mass, roughly consistent with the competing M87* mass measurement based on gas dynamics \citep{walsh13}. This is most likely a consequence of the choice of $r_{\mathrm{in}}\!=\!r_{\mathrm{ISCO}}\!=\!6M$, resulting in a larger image for the fixed object mass than in the case of a smaller $r_{\mathrm{in}}$. GRMHD simulations suggest that the ISCO has little importance for hot optically thin flows \citep{yuan14}, hence such a choice of $r_{\mathrm{in}}$ may be seen as inconsistent with the additional astrophysical or magnetohydrodynamical constraints. 
The models with spin $a = 0.8$ yield object mass consistent with the EHT and with the stellar dynamics measurement \citep{gebhardt11}. All models localize the maximum of the emission in the south of the image, see Fig. \ref{fig:data_fits}, right panel. The position angle of $200^\circ$, which approximates the expected orientation, given the observed position of the jet on the sky \citepalias[jet position angle in the observer's plane minus 90$^\circ$,][]{EHT5}, is indicated with a dashed horizontal line. We also see indication of counterclockwise rotation between first and last day of the observations, consistently with results reported in \citetalias{EHT4} and \citetalias{EHT6}.
None of the fits are of very high quality, as expected from very simple models with little number of degrees of freedom, see Figure \ref{fig:data_fits_quality}, and there is no clear indication of any of the models outperforming others. Some models, such as the ultracompact star with a surface located at 2.05$M$ (Fig. \ref{fig:disk_CompactStar}, middle column, fitting results not shown in this section) are in dramatic disagreement with the data, resulting in reduced $\chi^2$ errors larger than 100 for each of the EHT data sets. {Similarly, the model of a boson star shown in the last column of Fig. \ref{fig:disk_BS}, being rather symmetric in appearance, fits data poorly with reduced $\chi^2 > 30$ for all EHT data sets.} The fitting errors reported in Figure \ref{fig:data_fits_quality} are of similar magnitude as the average ones resulting from fitting individual snapshots from GRMHD simulations to the EHT data \citepalias{EHT5}. This supports the notion that geometric models could effectively represent mean properties of the more complicated simulations.

%---------------------------------------------------------------------
\section{Conclusion and perspectives}
\label{sec:conc}
%---------------------------------------------------------------------
%---------------------------------------------------------------------

In this article we develop a simple geometric model
for the inner accretion flow of M87*, the supermassive
black hole at the center of the galaxy M87. We use this
model to obtain predictions of the millimeter image of
the close surroundings of M87*, and compare them to the EHT findings. We have been focusing mainly on two questions. 

First, we tried to
develop on the recent studies devoted to improving
our understanding of the sharp highly-lensed features of
strong-field images~\citep{gralla19,Johnson2019}. Our findings regarding this issue are summarized in section~\ref{sec:conc_secondary} below. 

Secondly, we investigated whether
objects alternative to Kerr black holes (be they physically justified or not) could produce observational signatures similar to those seen by the EHT. {We have shown that interpreting EHT data sets with geometric flow models results in the image-domain morphology being consistent with the EHT findings (see also Figs. \ref{fig:bestfit_spin0}-\ref{fig:bestfit_spin08}), and several non-Kerr spacetimes fitting
  the EHT data similarly well as their Kerr counterparts.
  
  We showed that without an imposed assumption on the compact object mass, no spacetime curvature effects are needed to explain the current EHT results (see Michell-Laplace black hole). Even if a mass prior from stellar dynamics is assumed, exotic objects such as boson stars or Lamy wormholes can provide images consistent with the EHT observations of M87*{, with a favorable geometric configuration of the accretion flow}. Hence, we conclude that the published EHT observations, while consistent with the Kerr paradigm, do not provide a strong and unambiguous test of its validity.} 
Our results agree also with the
general point made earlier by~\citet{abramowicz02}.
Although, as stated by these authors, there cannot 
be any direct electromagnetic proof of the existence of
an event horizon, there is a hope for a robust detection of a
secondary ring by a more developed future version of the EHT. This would be of great importance, showing
the existence of null spherical orbits (which is however still
very different from showing that the object is a Kerr black hole). {On the other hand, resolving sharp, strongly lensed image features inconsistent with the ring geometry, such as what is seen for horizonless spacetimes with
photon orbits (see Appendix~\ref{app:BSLamy}), would make a very strong argument against the Kerr paradigm}.
Sections~\ref{sec:detect_secondary}
and~\ref{sec:constraining-epsilon} below are giving
our main conclusions regarding these topics. Finally, section~\ref{sec:perspective} gives some future research perspectives.

\subsection{Definition of the secondary ring}
\label{sec:conc_secondary}

Although the set of Kerr spherical orbits plays a crucial role in defining the thin bright ring region on the sky due to highly-lensed photons~\citep{teo03}, the properties of this feature depends strongly also on the proprieties of the accretion flow.

That is why we have defined in section~\ref{sec:kerrimages} the secondary ring
as the region on the observer's sky where the received null geodesics (i) have approached a Kerr spherical photon orbit within $\delta r \lesssim M$ in term of the radial Boyer-Lindquist coordinate $r$, and (ii) have visited the regions of the accretion flow emitting most of the radiation.

It is clear from this definition that the secondary ring is not only dictated
by gravitation. The astrophysics of the emitting gas has a role in
determining both the polar radius and azimuthal distribution of flux
in the ring. This dependency is of course at an extremely minute scale
($\approx \mu$as) and does not matter as far as the current EHT data are concerned. 
However, it will matter with future, higher-resolution space-VLBI data.

\subsection{Detecting the secondary ring}
\label{sec:detect_secondary}
\begin{figure}[htbp]
\centering
\includegraphics[width=\columnwidth]{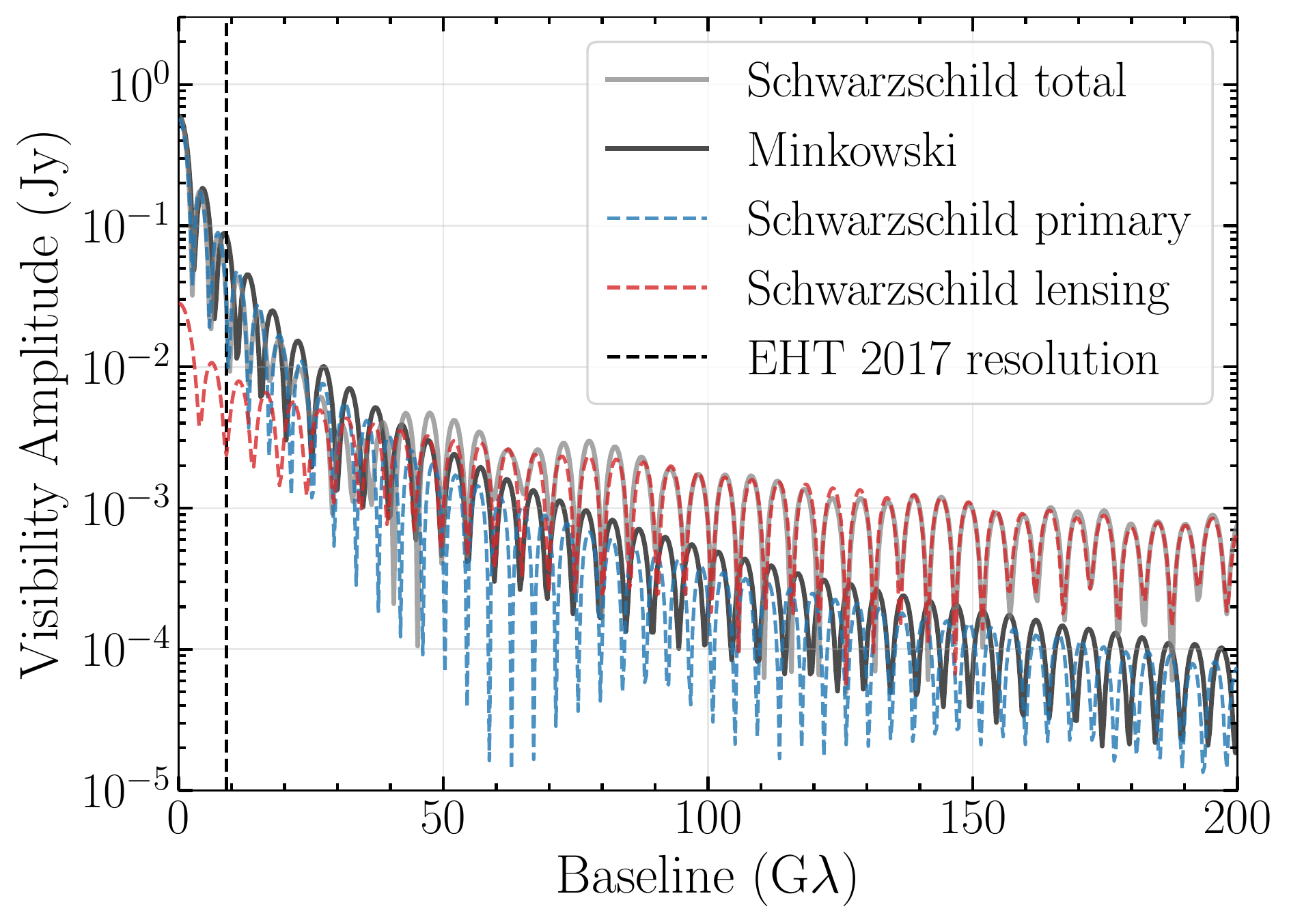}
\includegraphics[width=\columnwidth]{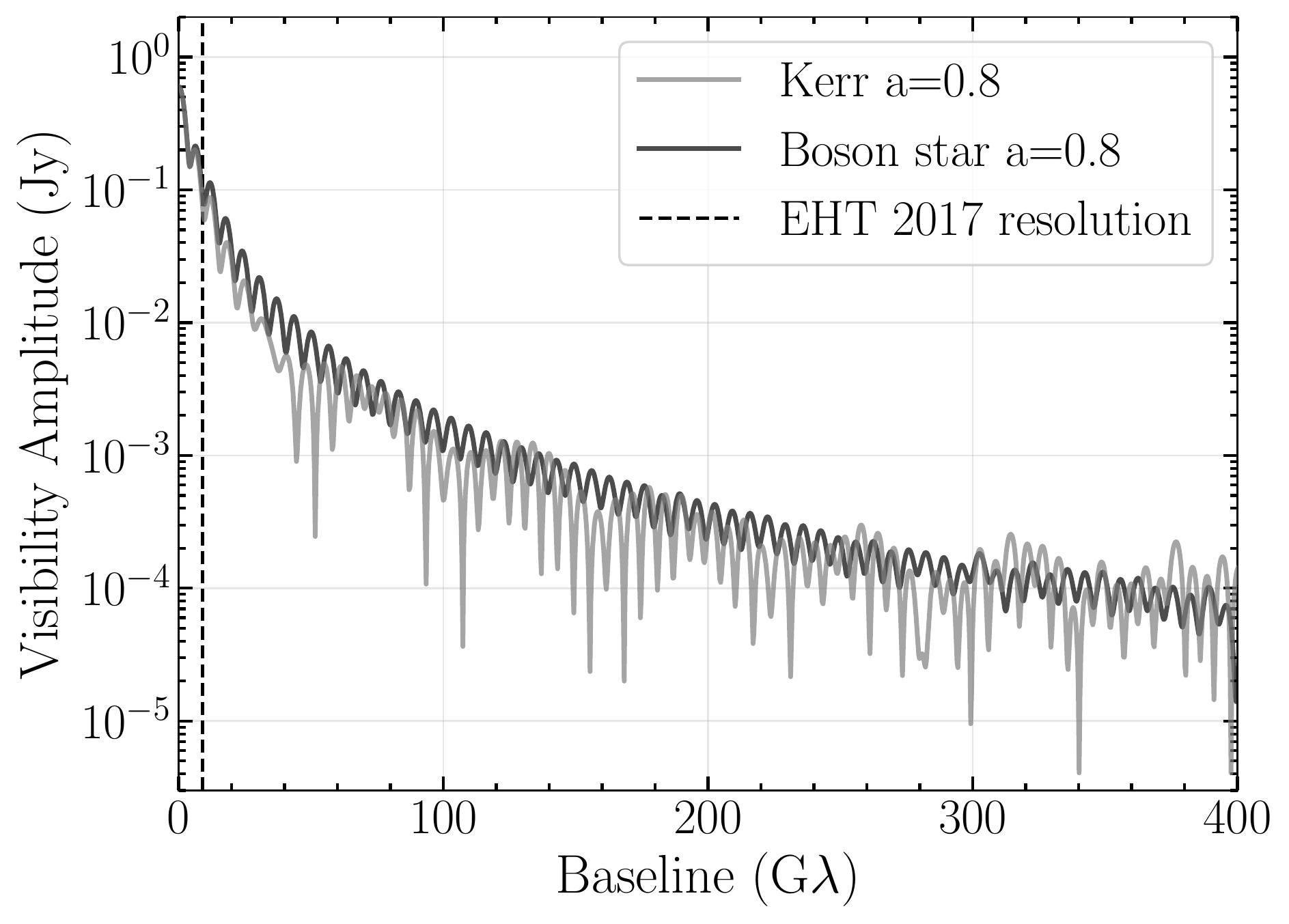}
\caption{\textit{Top:} 1D horizontal slice from a 2D Fourier transform of the images presented in Fig. \ref{fig:disk_minko}. For the Schwarzschild metric, the image has been decomposed into the primary image (smooth ring, dashed blue) and sharp secondary ring (dashed red). For very long baselines (large spatial frequencies) the secondary ring dominates the total emission. \textit{Bottom:} Similar, but for the images shown in Fig. \ref{fig:disk_BS}. While the boson star image does not exhibit a secondary ring, the amplitude fall-off is similar to that of the Kerr spacetime. 
} 
\label{fig:fourier}
\end{figure}

\citet{Johnson2019} proposed that VLBI observations with extremely long baselines could allow measuring the secondary ring properties to constrain the spin of M87*. This idea is based on the simple observation that the Fourier amplitudes of sharp image features decay slower with the spatial frequency than that of extended features, and therefore should dominate the signal on extremely long baselines. As an illustration, Fig. \ref{fig:fourier} (top) presents visibility amplitudes of images\footnote{1000$\times$1000 pixels images were used for this test.} of Schwarzschild black hole and Minkowski spacetime Michell-Laplace object (what is shown exactly is the horizontal slice through the amplitudes of the Fourier transforms of the two images). In this example, the contribution from a sharp ring feature clearly dominates the signal on baselines longer than 40 G$\lambda$, and at 100 G$\lambda$ the contribution of the sharp feature is one order of magnitude stronger than that of the primary image. Simply detecting an excess of power at very high spatial frequencies would therefore indicate the presence of a sharp feature. However, determining the exact character of the feature, e.g., photon ring, sharp edge, or Lamy wormhole's inner crescent, would likely require a more sophisticated modeling approach. This can be seen in Fig. \ref{fig:fourier} (bottom), where a very similar spectral-power fall-off characterizes both the Kerr spacetime, where a secondary ring is present, and the boson star spacetime, where that feature is missing. However, the two cases clearly differ in detailed structure. Detecting the presence of a sharp secondary ring feature would exclude solutions such as our Michell-Laplace relativistic black hole or a boson star without photon orbits, as that discussed in section \ref{sec:nonkerr}. Given the limitation of the Earth's size and atmospheric stability, space VLBI observation would be required in order to achieve this feat. Missions potentially capable of performing such a measurement are already being proposed, e.g., \cite{Millimetron,originWP}. 

\subsection{Constraining the ultracompact star surface location}
\label{sec:constraining-epsilon}

The constraints put on $\epsilon$ by the EHT images depend on the dynamic range of the images, that is on the ratio between the brightest image part and the least bright part, that can be reliably distinguished from the noise. Since the dynamic range of EHT images is of order of few tens, only an upper limit for
the contrast $\kappa$ between the brightest and
faintest parts of the image was given by \citetalias{EHT1}, $\kappa \ge 10$. Simulations show that the expansion of the EHT array planned for the 2020's decade should improve the dynamic range by an order of magnitude \citep{GroundWP}, tightening the constraints on $\epsilon$ by two orders of magnitude, see Eq. \ref{eq:epsilon}. These constraints will be ultimately limited by the jet emission from the region between the black hole and the observer, e.g., from the wall of the jet.
%---------------------------------------------------------------------
%---------------------------------------------------------------------
\subsection{Future perspectives}
\label{sec:perspective}

In this paper we have focused on comparisons between Kerr and non-Kerr spacetime models of M87*. {Hence, we did not present extensive studies of the influence of the model parameters on the image and its interpretation. As an example, we often fixed the radius $r_{\mathrm{in}}$ at the spacetime's ISCO. While the ISCO plays an important role in analytic models of relativistic {geometrically-thin} accretion as the inner disk radius \citep{nt73}, its relevance for {at least some realistic accretion scenarios} is expected to be less prominent \citep{abramowicz78,abramowicz10}. Since the ISCO radius strongly depends on spin, so does the size of the primary image of a disk model terminated at the ISCO -- which would result in spin-dependent diameter-mass calibration of the EHT results, contrary to the GRMHD predictions \citepalias{EHT5, EHT6}.  } %In a separate work (Wielgus et al., in prep) we will discuss the influence of model parameters on results, 
It will be interesting to discuss in the future the influence of the geometric model parameters on the obtained results, as well as to compare geometric models and GRMHD simulations, and their relevance for the interpretation of the M87* images.
%We will also analyze the impact of the black hole spin parameter on the observables.

We also plan to discuss the impact of an ejection flow on the observables in order
to be able to discuss all the possible states of M87* (disk- or jet-dominated). Our goal is to highlight the usefulness of the geometric models in order to
facilitate the understanding of the constraints that can be placed on the physical
parameters of the accretion flow and of the compact object.

\appendix
\section{Practical computation of the flux measured in the secondary ring}
\label{app:secondary}

The lensing and photon rings as defined by~\citet{gralla19}
are found by tracking the number of crossings of the equatorial plane of null 
geodesics.
In the realistic astrophysical context that we are dealing with here, such
a definition is not sufficient. Indeed, a geodesic can cross the equatorial plane
at a large radius, long after visiting the innermost regions of spacetime (as an example see the blue geodesic of
the lower panel of Fig.~\ref{fig:practicalsecondary}, which
crosses the equatorial plane at $r \approx 30M$).
Still, keeping track of the equatorial plane crossings
of geodesics is a very practical and easy-to-implement
way of dealing with highly-lensed geodesics, while
the general definition introduced in section~\ref{sec:kerrimages} is not very practical
to implement.
In order to determine the flux measured in the secondary
ring of the image,
we thus keep track of the null geodesics that cross the equatorial plane more than once
within a Boyer-Lindquist coordinate sphere $r < 10M$. 
This value is chosen
rather arbitrarily, after considering many highly bent geodesics as those illustrated
in Fig.~\ref{fig:explain_secondary}. This criterion is very easy to implement
in the context of our backward-ray-tracing code.

\begin{figure}[htbp]
\centering
\includegraphics[width=\columnwidth]{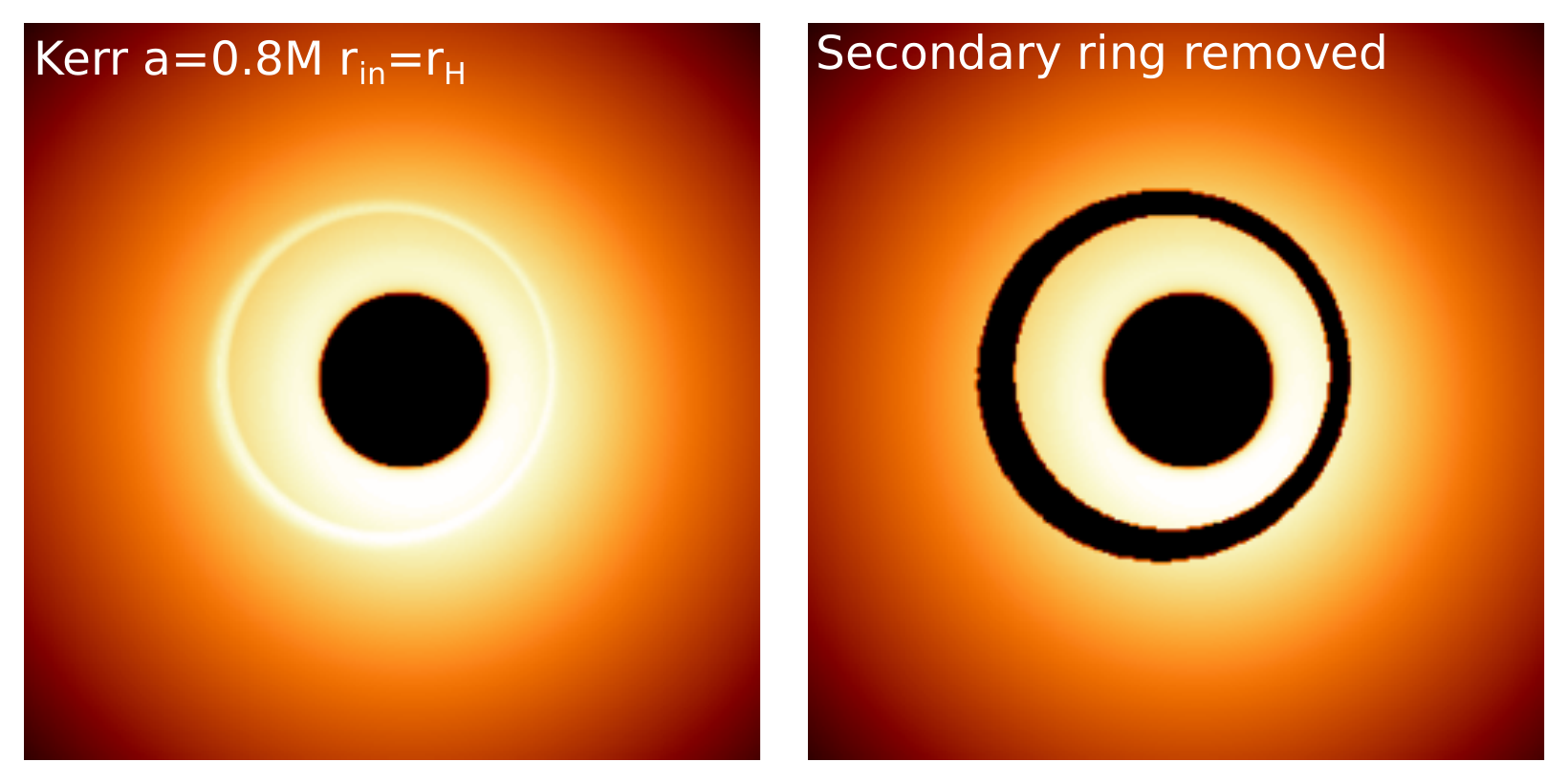}
\caption{\textit{Left:} An image of the central $80~\mu$as of a thick disk surrounding
a Kerr black hole with spin parameter $a=0.8M$, shown in logarithmic color scale.
\textit{Right:} same image, putting to zero all pixels corresponding
to geodesics that cross more than once the equatorial plane within the
coordinate sphere $r=10M$.
}
\label{fig:secondarycut}
\end{figure}

\begin{figure}[htbp]
\centering
\includegraphics[width=\columnwidth]{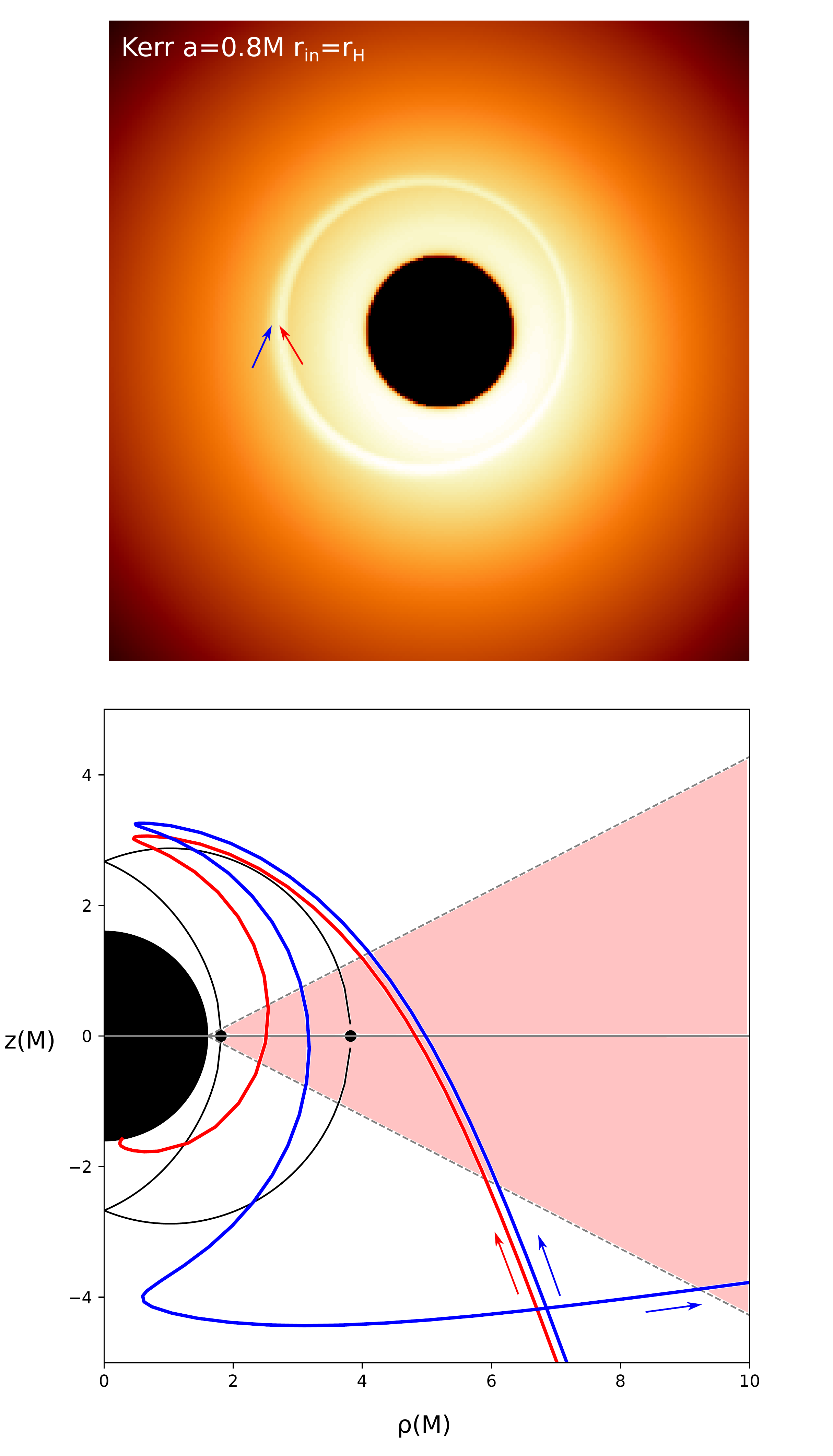}
\caption{Similar figure to Fig.~\ref{fig:explain_secondary} to which we refer for
the details of the lower panel. The upper panel is shown in logarithmic scale.
The red and blue geodesics of the lower panel correspond to the pixels labeled by the
red and blue arrows in the upper panel. The red geodesic is within the secondary ring
while the blue geodesic lies just outside. 
Note that the red geodesic asymptotically approaches
the event horizon when ray traced back in time.
} 
\label{fig:practicalsecondary}
\end{figure}

Fig.~\ref{fig:secondarycut} shows two images, with the secondary ring present
or removed by applying this recipe.
The cut seems by eye somewhat too large from this figure. The high-flux
pixels of the bright ring seem less extended than our mask. This is
due to the fact already highlighted above that the flux distribution in
the secondary ring depends a lot on the properties of the accretion flow.
Fig.~\ref{fig:practicalsecondary} illustrates this by following two geodesics, 
one of which (in red) belongs to the high-flux secondary ring, while the other 
(in blue) lies just outside. This figure shows that the red and blue geodesics
are extremely similar and cannot be simply distinguished based on pure gravitational
arguments such as number of crossings of the
equatorial plane, number of $\theta$ turning points, or
sharpness of these turning points. The red geodesic transports a lot of flux only because it visits the
innermost disk region, which is not the case for its blue counterpart.
This figure makes it clear that simple definitions of the secondary ring flux
like the one that we adopt are not enough to select the sharp high-flux region
of the image. The geometry of the accretion flow should also be taken into account,
as discussed in section~\ref{sec:kerrimages},
which makes a proper definition quite cumbersome. Note that it is not sufficient to
record the number of crossings of the thick disk, given that the blue geodesic
crosses it one more time than the red geodesic. 
Here, we have restricted ourselves
to the simple definition mentioned above, which allows to obtain reasonably
accurate estimates of the weight of the secondary ring flux.
However, this value is somewhat overestimated because
of the rather large angular thickness of the mask as compared to the thin secondary ring.

\section{Comparison between horizonless spacetimes with photon orbits}
\label{app:BSLamy}

\begin{figure*}[htbp]
\centering
\includegraphics[width=0.66\textwidth]{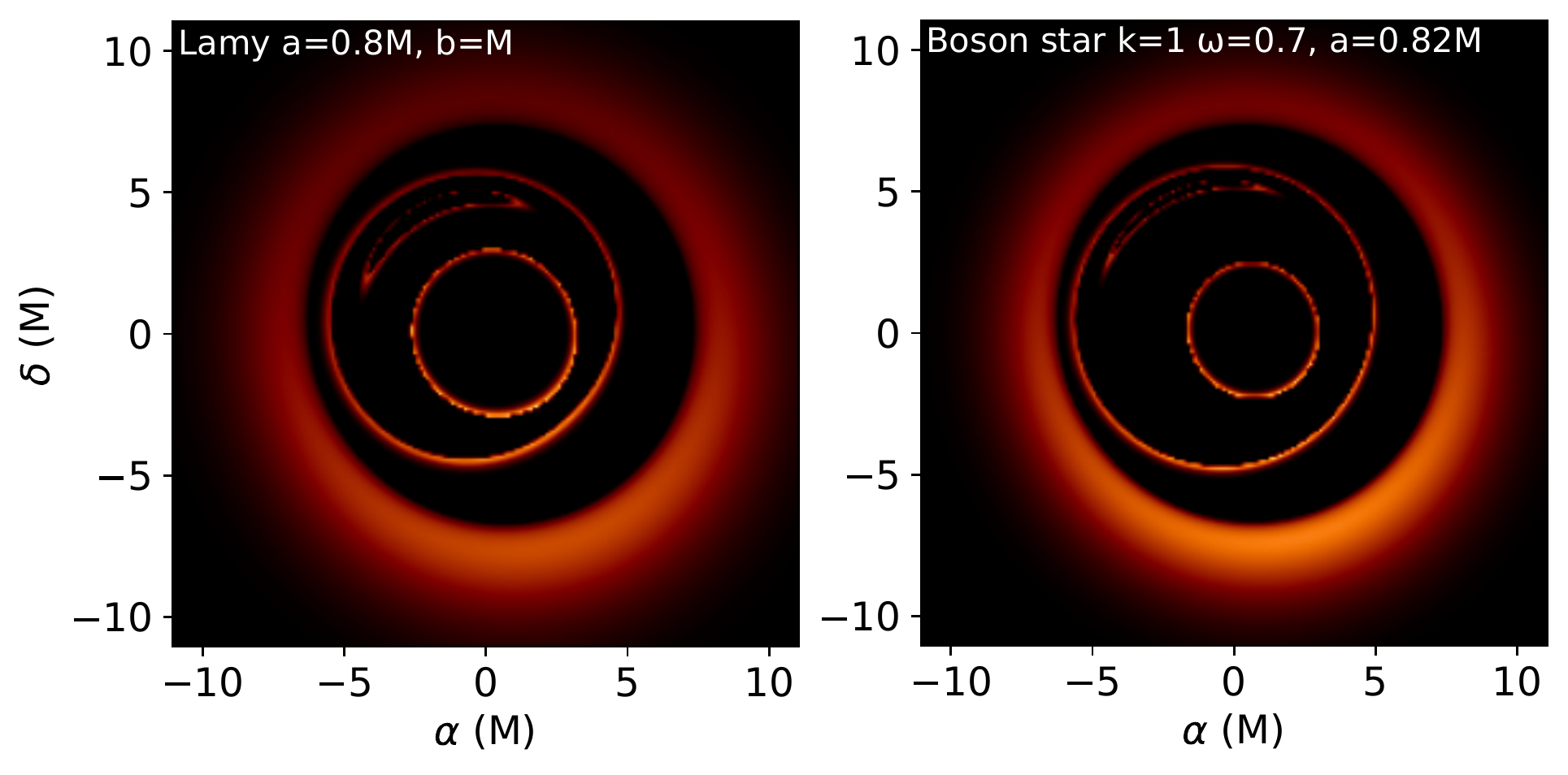}
\caption{Images on a field-of-view of $80 \, \mu$as of a thick disk with inner radius
$r_\mathrm{in} = 6M$ surrounding a \textbf{Lamy
wormhole} (left) or a \textbf{boson star} spacetime
with $k=1$ and $\omega=0.7$ (right). The 
compact object spin parameter is approximately
the same for both panels. The two spacetimes,
although extremely different at a theoretical
level,
are both horizonless and possess photon orbits.} 
\label{fig:BSLamy}
\end{figure*}

Section~\ref{sec:lamy}, and Fig.~\ref{fig:geod_compare} in particular, highlighted the similarity between our Lamy spacetime and a boson star spacetime
with k=1 and $\omega=0.7$. This comparison
was made on only one highly-bent null geodesic, which is extremely similar for
the two horizonless spacetimes.
Figure~\ref{fig:BSLamy} shows a comparison
of the full $80~\mu$as-field image between
these two spacetimes. The two images 
are strikingly similar.
In contrast, the ($k=1$, $\omega=0.77$)
boson star image is very different (see
top-middle panel of Fig.~\ref{fig:disk_BS},
computed with a field of view of $160\,\mu$as).
The important difference between the
two boson star spacetimes is the fact
that for ($k=1$, $\omega=0.77$), there is
no photon orbits, while for 
($k=1$, $\omega=0.7$), photon orbits 
exist~\citep[they appear at $\omega=0.75$,][]{cunha16}.
This is a good illustration that the main features of
highly-lensed images are dictated by
photon orbits~\citep[not only planar photon orbits, but the general set of fundamental photon orbits, that generalize Kerr spherical photon orbits, see][]{cunha17}. Even for spacetimes that
have nothing in common at a theoretical
level (like a Lamy wormhole and a boson star),
images are very similar as soon as 
similar photon orbits exist.
Lamy photon orbits have been studied
in detail
by~\citet{lamy18b}. 
Rotating-boson-star photon orbits have
been studied by~\citet{cunha16} and \citet{grandclement17}.
These works give the radius of the
unstable equatorial photon orbit,
which lies at $r_\mathrm{ph}=4.2M$
for the boson star spacetime
and $r_\mathrm{ph}\approx 2.2M$ for the
Lamy spacetime~\citep[we give an approximate value because in][a spacetime with $a=0.9M$ and $b=M$ is studied, thus slightly different from our case]{lamy18b}. The small difference
of scales in the two panels of Fig.~\ref{fig:BSLamy} are likely
due to this difference of location of the
equatorial photon orbit. A more detailed comparison would be necessary to analyze
further the similarity between these images
and link them to the properties of
fundamental photon orbits.
It is likely that a large class of horizonless spacetimes
with photon orbits will lead to images
similar to what we present here for
two particular spacetimes, {indicating sharp features that could be potentially resolved by the future VLBI arrays}.

\section{Best-fitting images}
\label{app:bestfit}
\begin{figure*}[htbp]
\centering
\includegraphics[width=\textwidth]{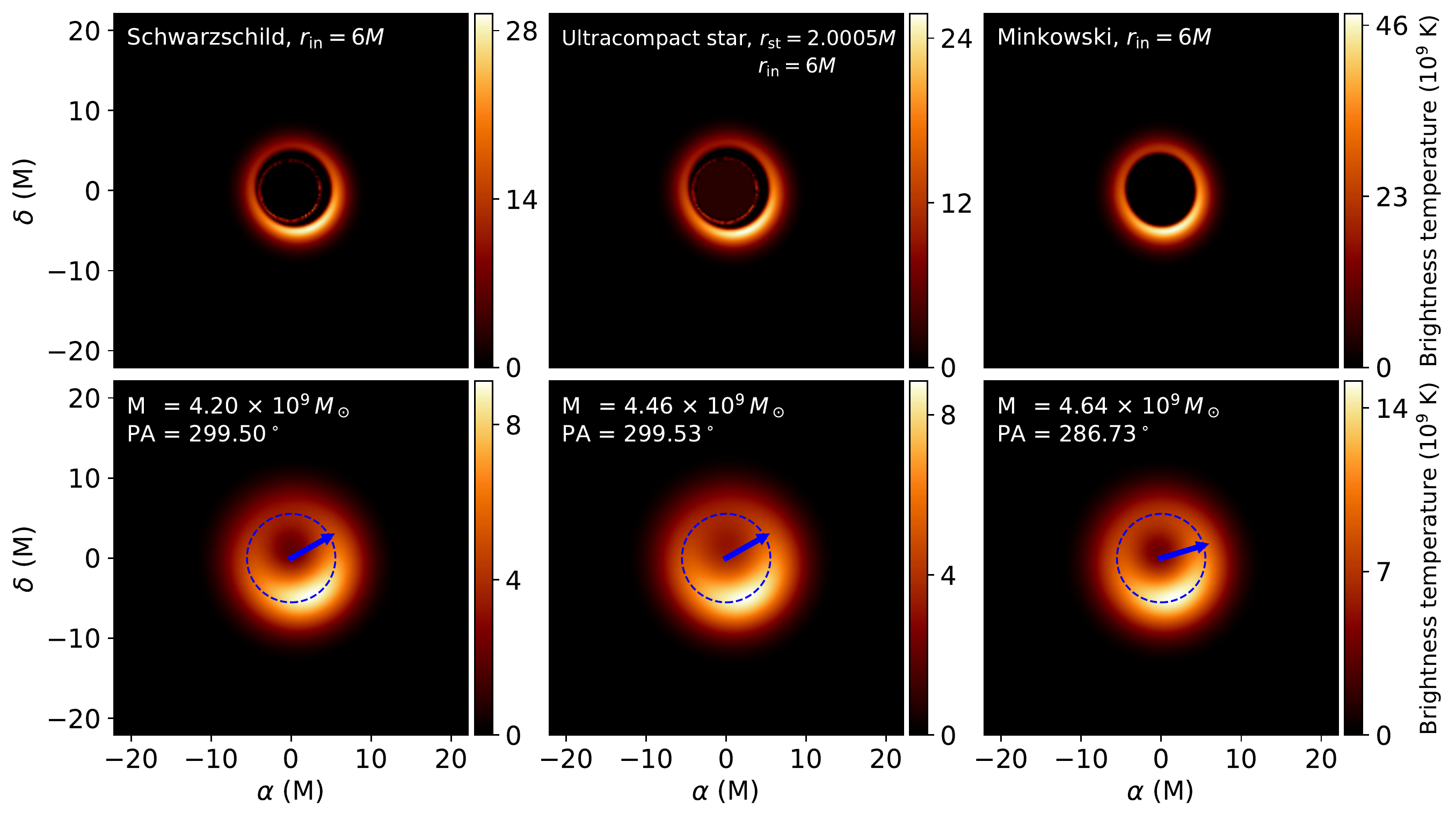}
\caption{Images best-fitting EHT data from April 11th 2017, HI band, corresponding to a geometrically thick accretion disk with inner radius
  $r_\mathrm{in} = 6M$ in a Schwarzschild spacetime (left column),
  in the spacetime of an {ultracompact star} with surface radius
  $r_\mathrm{st} = 2.0005M$ emitting blackbody radiation
  at the inner temperature of the accretion flow
  $T_\mathrm{e,in} = 8 \times 10^{10}$~K (middle column),
  or in a Minkowski spacetime (right column).
  As in all figures, the bottom row corresponds to the top row images blurred to the EHT resolution of $20\,\mu$as; the dashed blue circle has a diameter
  of $40\,\mu$as (size of the ring feature reported by the EHT) and the blue arrow shows the projected direction of the approaching jet. The best-fitting compact-object mass and jet position angle east of north are specified
   in each bottom panel.
} 
\label{fig:bestfit_spin0}
\end{figure*}

\begin{figure*}[htbp]
\centering
\includegraphics[width=\textwidth]{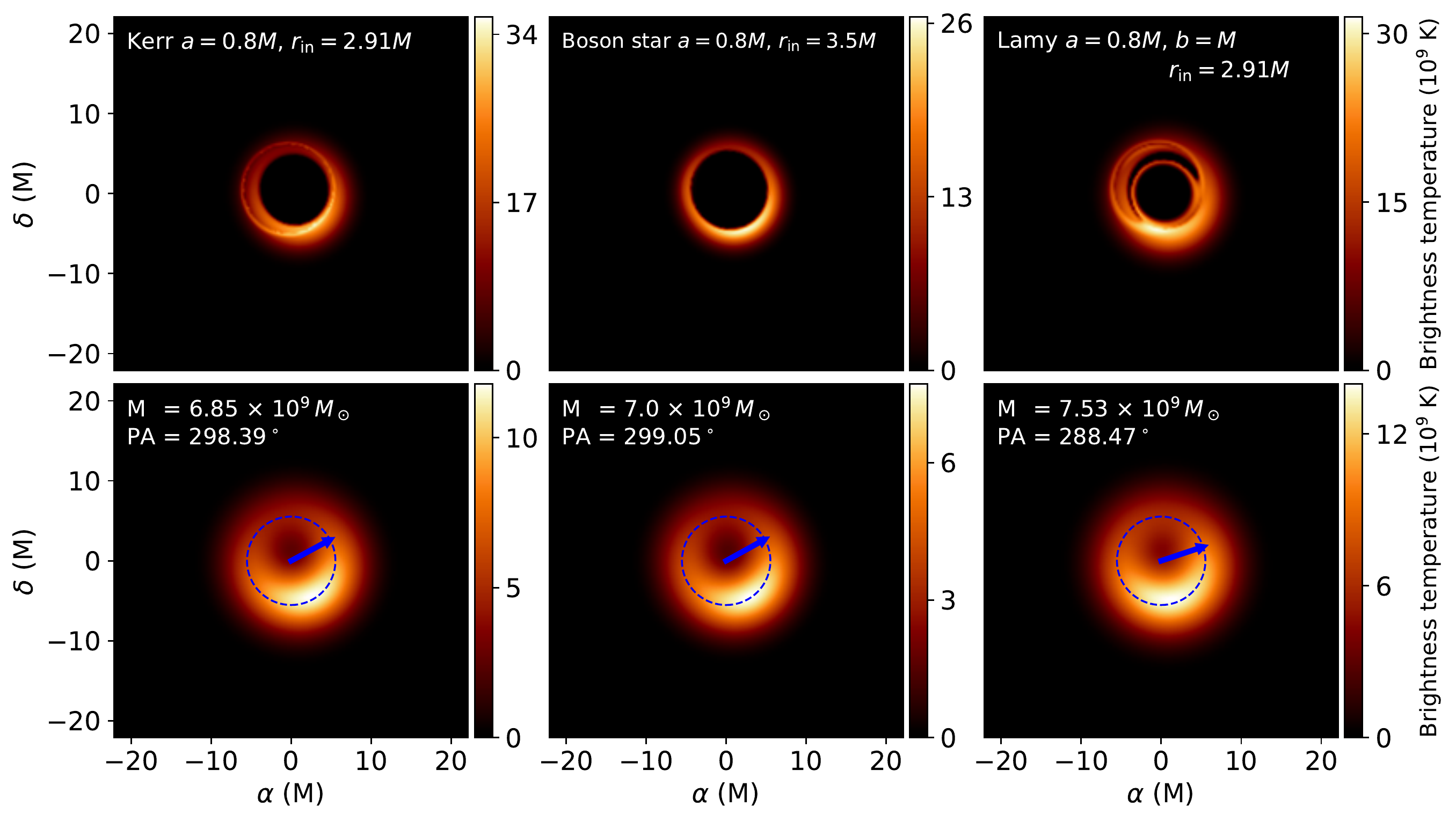}
\caption{Images best-fitting EHT data from April 11th 2017, HI band, corresponding to a geometrically thick accretion disk in a Kerr spacetime with spin
  $a=0.8M$ (left column),
  in a boson-star spacetime of the same spin
  with $k=1$ and $\omega=0.77$ (middle column),
  or in a Lamy spacetime of the same spin
  with $b=M$ (right column).
  The disk inner radius is of
  $r_\mathrm{in} = 2.91M$ for the left and right panels,
  and $r_\mathrm{in} = 3.5M$ for the central panel.
  As in all figures, the bottom row corresponds to the top row images blurred to the EHT resolution of $20\,\mu$as; the dashed blue circle has a diameter
  of $40\,\mu$as (size of the ring feature reported by the EHT) and the blue arrow shows the projected direction of the approaching jet. The best-fitting compact-object mass and jet position angle east of north are specified
   in each bottom panel.
} 
\label{fig:bestfit_spin08}
\end{figure*}

Figures~\ref{fig:bestfit_spin0}-\ref{fig:bestfit_spin08} present the results of the procedure of fitting models to the visibility domain data released by the EHT. In these examples we only show best fits to the data observed on April 11th 2017 in HI band {for several models that allowed to obtain fit quality similar as the Kerr spacetime examples}. Best-fitting jet position angle and object mass are given for the each type of object.

%---------------------------------------------------------------------
%---------------------------------------------------------------------
\section*{Acknowledgements}
{
We thank C. Gammie, R. Narayan, S. Doeleman, H. Olivares, and L. Rezzolla for helpful comments and discussions.
This work was supported by the Black Hole Initiative at Harvard University, which is funded by grants the John Templeton Foundation and the Gordon and Betty Moore Foundation to Harvard University.
JPL was supported by a grant from the French Space Agency CNES. EG acknowledges support from the CNRS program 80 PRIME TNENGRAV.
MAA acknowledges the Czech Science Foundation grant No.~17-16287S.
FHV acknowledges financial support from the Action
F\'ed\'eratrice PhyFOG of the Scientific Council of Observatoire de Paris.
  }
%---------------------------------------------------------------------
%---------------------------------------------------------------------

\bibliography{M87}
\bibliographystyle{aa}

\end{document}